\documentclass[nature,twocolumn,superscriptaddress,showkeys,nofootinbib]{revtex4-1}

\usepackage{graphicx} 
\usepackage{amsmath} 
\usepackage{epstopdf}
\usepackage[abs]{overpic} 
\usepackage{subfigure} 
\usepackage{cancel}
\usepackage{nicefrac}
\usepackage{amssymb}
\usepackage{xcolor}

\begin{document}

	\title{Spin-orbit coupling and Jahn-Teller effect in $T_d$ symmetry: an \textit{ab initio} study on the substitutional nickel defect in diamond}
	
	\author{Gerg\H{o} Thiering}  \email{thiering.gergo@wigner.hu} \affiliation{Wigner Research Centre for Physics, PO Box 49, H-1525, Budapest, Hungary}

	\author{Adam Gali} \email{gali.adam@wigner.hu} 
	\affiliation{Wigner Research Centre for Physics, PO Box 49, H-1525, Budapest, Hungary}
	\affiliation{Department of Atomic Physics, Institute of Physics, Budapest University of Technology and	Economics, M\H{u}egyetem rakpart 3., H-1111 Budapest, Hungary}

\begin{abstract}
			We analyze the spin-orbit and Jahn-Teller interactions in $T_d$ symmetry that are relevant for substitutional transition metal defects in semiconductors. We apply our theory to the substitutional nickel defect in diamond and compute the appropriate fine-leve structure and magneto-optical parameters by means of hybrid density functional theory. 
Our calculations confirm the intepretations of previous experimental findings that the 2.56-eV and 2.51-eV optical centres are associated with this defect. Our analysis of the electronic structure unravels possible mechanisms behind the observed optical transitions and the optically detected magnetic resonance signal, too. 
\end{abstract}


	\maketitle

\section{Introduction\label{sec:Intro}}
	
Nickel is a typical contaminant in high-temperature high-pressure synthesized (HPHT) diamonds that paved the way to observe various nickel related optical signals in diamond in the past decades~\cite{zaitsev2013optical, Nadolinny2017review}. However, various assignments of optical signals to defect structures were only tentative assumptions. Recent advances in the development of \textit{ab initio} methods made it feasible to revisit the optical centres with high prediction power and identify the microscopic structure of the optical centres. As a recent example, we have identified the nickel-vacancy complex by means of advanced density functional theory (DFT) methods~\cite{Thiering2021} as the origin of the 1.4-eV photo-luminescence (PL) centre~\cite{Dean_1965,Collins1983, Kupriyanov_1999} which was previously associated with the positively charged nickel interstitial~\cite{Isoya1990NIRIMg, Paslovsky1992, Pawlik1998, larico2009electronic}. Nickel-vacancy has $D_{3d}$ symmetry where group theory analysis combined with DFT calculations showed the electron-phonon coupling within Jahn-Teller (JT) effect formalism intertwins with spin-orbit coupling, and the full complexity of the system should be considered to obtain the key magneto-optical parameters for identification of nickel-related optical centres~\cite{Thiering2021}. In this paper, we apply this approach to the substitutional nickel defect (Ni$_\text{s}$) in diamond which has $T_d$ symmetry. Ni$_\text{s}$ introduces $d$ orbitals to the electronic structure which presumably also a subject of spin-orbit coupling, and partial occupation of degenerate $d$ states could lead to JT effect. As we shall show below, a very general group theory analysis of degenerate orbitals in $T_d$ symmetry is required for understanding this type of defect systems.

The Ni$_\text{s}$ configuration in diamond was observed in the electron paramagnetic resonance (EPR) spectra which is labelled as W8~\cite{LOUBSER1966firstW8, samoilovich1971electron, Isoya1990W8}. After the discovery of W8 EPR centre~\cite{LOUBSER1966firstW8}, the milestone experiments in the identification of Ni$_\text{s}$ were (i) the fingerprint of $^{61}\mathrm{Ni}$ isotope in the EPR spectrum with the hyperfine interaction between the electron spin and $I=3/2$ nuclear spin in isotopically $^{61}\mathrm{Ni}$ enriched HPHT diamond samples~\cite{samoilovich1971electron} and (ii) observation of $S=3/2$ electron spin with hyperfine signals of four identical $^{13}$C $I=1/2$ nuclear spins~\cite{Isoya1990W8}. The W8 EPR centre exhibits an isotropic $g=2.0310  \pm 0.0005$ tensor. These facts imply that the defect has $T_d$ symmetry. Furthmore, $S=3/2$ spin state indicates that it is a substitional defect in the negative charge state, i.e. Ni$_\text{s}^-$. This model was later confirmed by additional measurements and \textit{ab initio} calculations~\cite{Lowther1995, Larico2003, Larico2004, Larico2004basics, Larico2004Err, Chanier2013}. The ground state of Ni$_\text{s}^-$ is well understood: Ni$_\text{s}$ introduces a five times degenerate $d$ orbital of which level splits due to the tetrahedral crystal field of diamond as a triple degenerate one ($t_2$) lying in the fundamental band gap whereas the double degenerate $d$ orbital ($e$) is resonant with the valence band. The in-gap $t_2$ level is occupied by three electrons in the negative charge state which establishes the paramagnetic $S=3/2$ state. This electronic configuration is the $^4A_2$ orbital singlet multiplet state in $T_d$ symmetry.   

The correlation between the Ni-related optical centres and the W8 EPR centre were studied either indirectly such as common appearance in the appropriate optical and EPR spectra with similar estimated concentrations or directly via optically detected magnetic resonance (ODMR). 

In the former method, the correlation between the 1.883-eV and 2.51-eV Ni-related absorption centres and the W8 EPR centre was investigated. It was concluded that the 2.51-eV absorption centre is likely linked to the W8 EPR centre rather than the 1.883-eV absorption centre~\cite{Collins1998W82.51correlate}. Initial uniaxial stress measurements indicated that a triple degenerate $T_2$ state is involved in the 2.51-eV absorption spectrum~\cite{Nazar1994T2T} which was later revisited and concluded that ${}^4 A_2 \leftrightarrow{}^4 T_2$ optical transition is involved in the absorption process~\cite{Nazar2001W82.51A1T2}, being consistent with the ground state of Ni$_\text{s}^-$. Although, the $^4T_2$ state is not optically allowed from $^4A_2$ ground state but it was assumed that spin-orbit coupling between the $^4T_2$ and ${}^4 A_2$ makes this transition partially allowed. The shape of the first peak exhibit a doublet feature separated by 1.5~meV~\cite{Nazar2001W82.51A1T2}. They interpreted this feature to the spin-orbit splitting of the $^4T_2$ state. Beside this feature a replica shows up at around 16.5~meV above the zero-phonon-line (ZPL) energy which is broadened compared to the shape of the ZPL emission~\cite{Nazar2001W82.51A1T2}. The origin of the replica was not explained.

In the latter method, the 2.56-eV PL centre could be directly linked to the W8 EPR centre via ODMR measurements at cryogenic temperatures~\cite{Pereira1994, Nazare1995}. Two individual features were observed in the ODMR spectrum excited by 325-nm laser in a 35-GHz microwave cavity~\cite{Nazare1995}: (i) an electron spin resonance with the isotropic $g=2.0324(5)$ associated with the W8 EPR centre with producing a PL spectrum in the ODMR contrast which agrees well with the 2.56-eV PL spectrum in terms of the zero-phonon-line (ZPL) energy and the features in its phonon sideband; (ii) the hyperfine splitting originating from the $^{14}\mathrm{N}$ $I=1$ nuclear spin in the P1 EPR centre, i.e. the neutral substitutional nitrogen donor, N$_\text{s}^0$ with $g\approx2.00$ which produces a very broad gaussian-shape PL spectrum in the ODMR contrast. The ODMR contrast of the $S=1/2$ P1 centre was explained by a donor-acceptor pair (DAP) model, where the unknown acceptor $A$ interacts with the N$_\text{s}$ donor as
\begin{equation}
N_\text{s}^0 + A^0 \rightarrow N_\text{s}^+ + A^- + h\nu \text{,}
\end{equation}
where $h \nu$ photon is emitted for which the energy depends the distance between the N$_\text{s}$ and the acceptor $A$, so it results in a broad fluorescence spectrum for the ensemble of DAPs. If the unknown acceptor is boron (acceptor level is at 0.37~eV above the valence band maximum, VBM) then the fluorescence spectrum can be well explained by taking the huge reconstruction energy of the N$_\text{s}$ upon ionisation into account~\cite{Nazare1995}. It was speculated that electron spin resonance of the acceptor in the ODMR spectrum was not observed because if the acceptor state is an effective-mass-like, then the degeneracy of the top of the VBM causing its EPR signal to be strongly strain broadened and difficult to detect, as in all cubic semiconductors~\cite{Nazare1995}. Regarding the Ni-related ODMR feature, it is more difficult to come up with a unique model that explains the observation of the ground state EPR of Ni, in the luminescence. Nazar\'e \textit{et al.} speculated~\cite{Nazare1995} that one possibility is spin-dependent hole transfer from a distant acceptor to Ni$_\text{s}^-$, producing an excited luminescent state of Ni$_\text{s}^0$. We note that this may explain that the observed isotropic $g$-factor in ODMR spectrum (2.0324) slightly differs from the $g$-factor in the W8 EPR spectrum (2.0310)~\cite{Nazare1995}. It is worth to note too that the 2.56-eV PL centre can be activated though a broad excitation at $200\pm20$~nm range ($\sim6.2\pm0.6$~eV) efficiently as performed by photo-luminescence excitation (PLE) measurements~\cite{Lu2015}. This suggests that carriers trapping is indeed involved behind the ODMR process of Ni$_\text{s}$ defect. We note that the reported negative ODMR contrast of the 2.56-eV PL in Ref.~\onlinecite{Pereira1994} was about $10^{-5}$ in their experimental conditions (excitation wavelength of 365~nm applied in 36.2-GHz microwave cavity) which showed up as a positive contrast without reporting its magnitude in Ref.~\onlinecite{Nazare1995} in their experimental conditions (excitation wavelength of 325~nm applied in 35-GHz microwave cavity). This fact also implies that the ODMR process is not intrinsic to the Ni$_\text{s}$ defect alone. Photo-EPR measurements were carried out in nickel doped diamonds and found that the intensity of W8 EPR signal decreases by photo-excitation with $\sim2.5$~eV threshold energy which can be reinstated by photo-excitation with $\sim3.0$~eV~\cite{Pereira2003}.

\section{Methodology\label{sec:Methods}}

We characterise Ni$_\text{s}$ in diamond by plane wave supercell calculations within spin-polarised DFT as implemented in the \textsc{vasp} 5.4.1 code~\cite{Kresse:PRB1996}. We determine the electronic structure within the Born-Oppenheimer approximation where the ions are treated as classical particles. We relaxed the atomic positions until the forces acting on the ions fall below 10$^{-2}$~eV/\AA. We embed the defect in a 512-atom diamond supercell and we sample the Brillouin-zone at the $\Gamma$-point. We used the projector-augmentation-wave-method (PAW)~\cite{Blochl:PRB1994, Blochl:PRB2000} as implemented in \textsc{vasp}. We used the standard PAW projector for carbon and Ni\_pv PAW projector for nickel that includes the $3p$ atomic orbitals as valence. We applied a plane wave cut-off energy at 370~eV for the plane wave basis that was already proven to be convergent for SiV defect and many other defect system in diamond~\cite{Gali:PRB2013, Thiering2015, thiering2018theory,  Thiering2021, Thiering2017SOC, Thiering2016, thiering2018emph}.  We calculate the excited states with the constrained-occupation DFT method or $\Delta$SCF method~\cite{Gali2009}. We employed the Heyd-Scuseria-Ernzerhof (HSE06) hybrid functional~\cite{Heyd03, Krukau06} which reproduces the experimental band gap and the charge transition levels in diamond or other group-IV semiconductors within 0.1~eV accuracy~\cite{Deak:PRB2010, Gali2009}. For the charged supercell we applied the Freysoldt-Neugebauer-van de Walle correction to the total energy~\cite{Freysoldt2009}.
%
%
We use our home-built code to compute the electron-phonon spectrum of the Jahn-Teller modells that we successfully applied to other systems too~\cite{Thiering2017SOC, thiering2018emph}.

\section{Results\label{sec:Results and Discussion}}
	
We combine group theory description of the Ni$_\text{s}$ defect with DFT calculations. We start with the calculated electronic structure and charge transition levels of the defect. We identify the most common charge states under typical experimental conditions in this section. We continue with the detailed discussion of the ground state manifolds of the most relevant charged states of the defect and then we describe the optically accessible excited states. Finally, we discuss our findings in the light of the observed 2.51-eV absorption centre and the 2.56-eV PL centre including the possible mechanisms behind the ODMR signature. 
	
\subsection{Electronic structure of Ni$_\text{s}$ \label{sec:KS}}
	
\begin{figure*}[] 
		\includegraphics[width=\textwidth]{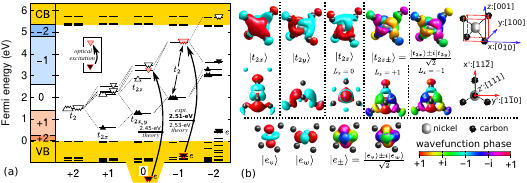} 
  	\caption{\label{fig:NisKS}(a) Single particle levels and charge transition levels for Ni$_\text{s}$. Note that the spin-polarisation in the applied method splits the occupied ($\blacktriangle\text{,}\blacktriangledown$) and unoccupied ($\vartriangle$,$\triangledown$) defect levels,
 and the Fock-exchange potential in the hybrid DFT further lifts the degeneracy within the same spin channel for the occupied and unoccupied orbitals. Additionally, we show the lowest possible optical excitation between $t_2$ and $e$ orbitals for Ni$_\text{s}^{0}$ and Ni$_\text{s}^{-}$. We note that the $e$ orbitals Ni$_\text{s}^{0}$ are smeared into multiple valence band states, thus its position is only approximate. However, upon $\Delta$SCF procedure the $e$ hole is getting localized and elevated into the band gap. (b) Kohn-Sham orbitals of Ni$_\text{s}^-$ in diamond. In the real wavefunctions, the red and cyan isovalent surfaces correspond to positive and negative values, respectively. We plot the $|t_{2x,2y,2z}\rangle$ orbitals from Ni$_\text{s}^-$'s triply degenerate ($\blacktriangle\!\!\blacktriangle\!\!\blacktriangle$) occupied $t_2$ levels. The $|e_{u,w}\rangle$ are taken from Ni$_\text{s}^{2-}$'s nearly doubly degenerate ($\blacktriangledown\!\!\blacktriangledown$) occupied $e$ orbitals. }

\end{figure*}

The Ni atom introduces $3d$ orbitals which splits to double degenerate $e$ orbitals and triple degenerate $t_2$ orbitals in the tetrahedral crystal field of diamond. In the defect molecule diagram, these $3d$ orbitals and the $4s$ orbitals interact with the four dangling bonds of the neighbour carbon atoms that results in $a_1$ and $t_2$ molecular orbitals under $T_d$ symmetry. The $4s$ and $a_1$ orbitals can form bonding and anti-bonding molecular orbitals lying deep in the valence band and high in the conduction band, respectively. The $e$ orbitals do not recombine with the dangling bonds and they remain atomic like. The level of $e$ orbitals is resonant with the valence band in the neutral charge state of the defect as plotted in Fig.~\ref{fig:NisKS}(a). Finally, the $t_2$ representations of the $3d$ orbitals and the dangling bonds again form bonding and antibonding molecular orbitals. The bonding combination falls deep in the valence band whereas the antibonding combination appears in the gap. In the context, we call this antibonding $t_2$ orbital simply the $t_2$ state in the gap which is responsible for the ground state manifolds of the Ni$_\text{s}$ defect. In the optical excitation, the atomic like $e$ states may play a role. The calculated Kohn-Sham states are depicted in Fig.~\ref{fig:NisKS}(b).   

According to this defect molecule analysis, the defect has the $a_{1}^{2}t_{2}^6e^{4}t_{2}^{2-q}$ electronic configuration, where $q$ is the charge state of the defect. As the triple degenerate $t_2$ may accommodate six electrons, the possible charge states may go from the $(2+)$ to $(4-)$. We calculated the total energies of the corresponding charge states ($E^q_\text{tot}$) and calculated the adiabatic charge transition levels as $(q|q-1)= E^q_\text{tot} - E^{q-1}_\text{tot} + q E_\text{VBM}$, where $E_\text{VBM}$ is the calculated VBM energy and now it is aligned to zero in Fig.~\ref{fig:NisKS}(a).	

We find that the aniticipated $(3-)$ and $(4-)$ states are not stable at all, and the $(2+)$ and $(2-)$ charge states are marginally stable as a function of the Fermi-level in diamond. Therefore, we do not analyze the $(2+)$ and $(2-)$ charge states further. The calculated $(0|-)$ level is at about $E_\text{VBM}+2.6$~eV, which means that it is about $E_\text{CBM}-2.8$~eV, where $E_\text{CBM}$ is the energy of the conduction band minimum (CBM). In other words, if the defect was in the neutral charge state then about 2.6~eV photo-excitation energy is required to ionise to $(-)$ charge state, and if the defect was in the $(-)$ charge state then about $2.8$~eV photo-excitation energy is required to re-ionise it back to the neutral charge state. Ni$_\text{s}^+$ can be readily ionised to the neutral charge state by the photo-excitation threshold energy at about 1.5~eV. It can be read out from Fig.~\ref{fig:NisKS}(a) that Ni$_\text{s}^+$ may be stable in $p$-type diamond and near-infrared photo-excitation is needed below the photo-ionisation threshold energy in order to observe its absorption or fluorescence signal. In practice, these conditions are not fulfilled, thus we will only briefly discuss the ground state of Ni$_\text{s}^+$. We rather focus our discussion to the negatively charged and neutral Ni$_\text{s}$ defects.	

We briefly discuss now the 2.51-eV absorption and 2.56-eV PL centres in the light of the calculated photo-ionisation threshold energies. According to the calculated $(0|-)$ level at around $E_\text{VBM}+2.6$~eV, both optical centres might correspond to the neutral excitation of both Ni$_\text{s}^0$ and Ni$_\text{s}^-$ defects. The 2.56-eV PL centre has the ZPL energy just below the calculated Ni$_\text{s}^0$$\rightarrow$Ni$_\text{s}^-$ ionisation threshold energy. Thus, the calculated ionisation threshold energies could not distinguish alone which optical centre belongs to which defect charge state. In order to identify these optical centres, we analyze the ground and excited states of Ni$_\text{s}^-$ and Ni$_\text{s}^0$ in detail in the next sections. 

It is intriguing that the 2.56-eV emission can be observed above-band-gap excitations~\cite{Lu2015}, and its ODMR signals were observed by 325-nm (3.81-eV) photo-excitation~\cite{Nazare1995} and 365-nm (3.40-eV) photo-excitation~\cite{Pereira1994} experiments. In the latter two experiments, the photo-excitation energies are both above the ionisation threshold energies for both the Ni$_\text{s}^0$$\rightarrow$Ni$_\text{s}^-$ and Ni$_\text{s}^-$$\rightarrow$Ni$_\text{s}^0$ processes, where the electron is promoted from the valence band to the empty defect level and from the occupied defect level to the conduction band, respectively. The calculated $(0|-)$ level at around $E_\text{VBM}+2.6$~eV implies that the electron is promoted from a deeper energy region of the valence band in the first process rather than it is excited from the occupied defect level to little above the conduction band minimum in the second process. As the electron density of states rapidly increases going farther from the band edges this implies that Ni$_\text{s}^0$$\rightarrow$Ni$_\text{s}^-$ process is much more efficient than Ni$_\text{s}^-$$\rightarrow$Ni$_\text{s}^0$ process. As a conseuqence, the ultraviolet excitation will stabilise the Ni$_\text{s}^-$ charge state, so the defect stays much longer time in the $(-)$ charge state than in the $(0)$ charge state.
	
\subsection{Ground state of Ni$_\text{s}^-$ \label{sec:NisGND}}

It is likely that ultraviolet photo-excitation stabilises the negative charge state of Ni$_\text{s}$, and the totally filled $t_2$ electronic configuration of Ni$_\text{s}^-$ can be readily analyzed as a basis for the more complex electronic configurations. Therefore, we start the analysis with the ground state electronic structure of Ni$_\text{s}^-$.

According to first Hund's rule the three electrons in the triple degenerate $t_2$ orbitals will form a high-spin $S=3/2$ electronic configuration. The wavefunction of such quartet state can be expressed as
	\begin{equation}
	\label{Nisneg_WAVE}
	|^{4}A_{2}\rangle=\underset{\mathcal{A}|xyz\rangle}{\underbrace{\mathcal{A}|t_{2x}t_{2y}t_{2z}\rangle}}\otimes\left\{ \begin{array}{c}
	\left|\uparrow\uparrow\uparrow\right\rangle \\
	\mathcal{S}\left|\uparrow\uparrow\downarrow\right\rangle \\
	\mathcal{S}\left|\uparrow\downarrow\downarrow\right\rangle \\
	\left|\downarrow\downarrow\downarrow\right\rangle 
	\end{array}\right.
	\text{,}
	\end{equation} 
	where we define the three particle ($\mathcal{S}$) symmetrisation  
	\begin{equation}
	\label{Nisneg_WAVE_spinsym}
	\mathcal{S}\left|\uparrow\uparrow\downarrow\right\rangle =\frac{1}{\sqrt{3}}\left(\left|\uparrow\uparrow\downarrow\right\rangle +\left|\uparrow\downarrow\uparrow\right\rangle +\left|\downarrow\uparrow\uparrow\right\rangle \right)
	\end{equation} 
	and ($\mathcal{A}$) anti-symmetrisation 
	\begin{equation}
	\label{Nisneg_WAVE_orbitalsym}
	\begin{aligned}\mathcal{A}|t_{2x}t_{2y}t_{2z}\rangle=\mathcal{A}|xyz\rangle=\\
	\frac{1}{\sqrt{6}}\left(|xyz\rangle-|xzy\rangle+|yzx\rangle-|yxz\rangle+|zyx\rangle-|zxy\rangle\right)
	\end{aligned}
	\end{equation} 
	operators for the sake of simplicity. In addition, we use shortened $(x,y,z)$ notation to label $(t_{2x},t_{2y},t_{2z})$ orbitals in the context from now on. Our results are in agreement with previous \textit{ab initio} studies~\cite{Larico2004basics, Chanier2013}, the ground state is a $|^{4}A_{2}\rangle$ multiplet. The $S=3/2$ spin of W8 EPR centre emerges from the three unpaired electrons fillling the $t_{2}$ orbitals in Ni$_\text{s}^-$.

\subsection{Shelving doublet states of $t_{2}^{(3)}$ electronic configurations in Ni$_\text{s}^-$ \label{sec:NisGNDdoulets}}
	
It is clear that $t_{2}^{(3)}$ electronic configuration can form doublet states too as the total number of possible combinations of muliplet states is $\binom{6}{3}=20$ where $^4A_2$ quartet state produces only four multiplet states from them. The other configurations form doublet manifolds that should be expressed as a combination of Slater-determinants. First, we determine the possible configurations with single Slater-determinants. There are 8 possible occupations where we pin electrons on individual $x,y,z$ spin-orbitals as
	\begin{equation}
			\label{Nisgnddimensionality}
	\begin{array}{c}
	\mathcal{A}\bigl|x_{\uparrow}y_{\uparrow}z_{\uparrow}\bigr\rangle\,,\,\mathcal{A}\bigl|x_{\uparrow}y_{\uparrow}z_{\downarrow}\bigr\rangle\,\text{,}\,\mathcal{A}\bigl|x_{\uparrow}y_{\downarrow}z_{\uparrow}\bigr\rangle\,,\,\mathcal{A}\bigl|x_{\downarrow}y_{\uparrow}z_{\uparrow}\bigr\rangle\\
	\mathcal{A}\bigl|x_{\downarrow}y_{\downarrow}z_{\downarrow}\bigr\rangle\,\text{,}\,\mathcal{A}\bigl|x_{\downarrow}y_{\downarrow}z_{\uparrow}\bigr\rangle\,\text{,}\,\mathcal{A}\bigl|x_{\downarrow}y_{\uparrow}z_{\downarrow}\bigr\rangle\,,\,\mathcal{A}\bigl|x_{\uparrow}y_{\downarrow}z_{\downarrow}\bigr\rangle
	\end{array}\:\text{,}
	\end{equation} 
	where $S_{z}=\pm\frac{3}{2}$ combinations and the totally symmetric $S_{z}=\pm\frac{1}{2}$ combinations are reserved for the $|^{4}A_{2}\rangle$
	ground state as
	\begin{equation}
	\label{NisgndA2}
	|^{4}A_{2}\rangle=\!\!\left\{ \begin{array}{c}
	\mathcal{A}\bigl|x_{\uparrow}y_{\uparrow}z_{\uparrow}\bigr\rangle\\
	\!\!\frac{1}{\sqrt{3}}\Bigl(\mathcal{A}\bigl|x_{\uparrow}y_{\uparrow}z_{\downarrow}\bigr\rangle+\mathcal{A}\bigl|x_{\uparrow}y_{\downarrow}z_{\uparrow}\bigr\rangle+\mathcal{A}\bigl|x_{\downarrow}y_{\uparrow}z_{\uparrow}\bigr\rangle\Bigr)\!\!\\
	\!\!\frac{1}{\sqrt{3}}\Bigl(\mathcal{A}\bigl|x_{\downarrow}y_{\downarrow}z_{\uparrow}\bigr\rangle+\mathcal{A}\bigl|x_{\downarrow}y_{\uparrow}z_{\downarrow}\bigr\rangle+\mathcal{A}\bigl|x_{\uparrow}y_{\downarrow}z_{\downarrow}\bigr\rangle\Bigr)\!\!\\
	\mathcal{A}\bigl|x_{\downarrow}y_{\downarrow}z_{\downarrow}\bigr\rangle
	\end{array}\right\}   \text{,}
	\end{equation} 	
	that is equivalent to Eq.~\eqref{Nisneg_WAVE}. Therefore, the remaining four-dimensional subspace spans a $|^{2}E\rangle$ state as
		\begin{equation}
		\label{NisgndE}
	|^{2}E\rangle=\!\!\left\{ \begin{array}{c}
	\!\!\frac{1}{\sqrt{6}}\Bigl(2\mathcal{A}\bigl|x_{\uparrow}y_{\uparrow}z_{\downarrow}\bigr\rangle-\mathcal{A}\bigl|x_{\uparrow}y_{\downarrow}z_{\uparrow}\bigr\rangle-\mathcal{A}\bigl|x_{\downarrow}y_{\uparrow}z_{\uparrow}\bigr\rangle\Bigr)\!\!\\
	\frac{1}{\sqrt{2}}\Bigl(\mathcal{A}\bigl|x_{\downarrow}y_{\uparrow}z_{\downarrow}\bigr\rangle-\mathcal{A}\bigl|x_{\downarrow}y_{\downarrow}z_{\uparrow}\bigr\rangle\Bigr)\\
	\!\!\frac{1}{\sqrt{6}}\Bigl(2\mathcal{A}\bigl|x_{\downarrow}y_{\downarrow}z_{\uparrow}\bigr\rangle-\mathcal{A}\bigl|x_{\downarrow}y_{\uparrow}z_{\downarrow}\bigr\rangle-\mathcal{A}\bigl|x_{\uparrow}y_{\downarrow}z_{\downarrow}\bigr\rangle\Bigr)\!\!\\
	\frac{1}{\sqrt{2}}\Bigl(\mathcal{A}\bigl|x_{\uparrow}y_{\downarrow}z_{\uparrow}\bigr\rangle-\mathcal{A}\bigl|x_{\uparrow}y_{\uparrow}z_{\downarrow}\bigr\rangle\Bigr)
	\end{array}\right\}  \text{.}
	\end{equation}

By constraint occupation of the $t_2$ Kohn-Sham orbitals, the total energy of the system can be calculated by spin-polarised HSE06 functional as given in Eq.~\eqref{Nisgnddimensionality}. The exchange-correlation functional of HSE06 may not be able to capture the high correlation which can be described as combinations of Slater-determinants and the resulting solutions will be not the true eigenstate of the doublet spin state.

With this caveat, we carried out \textit{ab initio} calculations for the single Slater-determinant configuration, e.g. $\mathcal{A}\bigl|x_{\uparrow}y_{\uparrow}z_{\downarrow}\bigr\rangle$ of the Kohn-Sham $t_2$ states which yields 0.35~eV with respect to the calculated total energy of $^4A_2$ state. This was achieved by restricting the symmetry to $T_d$ symmetry. By enabling the symmetry reduction for the $\mathcal{A}\bigl|x_{\uparrow}y_{\uparrow}z_{\downarrow}\bigr\rangle$ electronic configuration, the energy gain has become negligible at 1.4~meV. We conclude that there will be no Jahn-Teller effect for the doublet configurations in the first order. The $\ensuremath{\mathcal{A}\bigl|x_{\uparrow}y_{\uparrow}z_{\downarrow}\bigr\rangle}$	determinant is an 33\% admixture of $|^{4}A_{2}\rangle$ and 66\% of
	$|^{2}E\rangle$ multiplets, thus we expect the level of $|^{2}E\rangle$ lying above that of $|^{4}A_{2}\rangle$ by roughly 0.35~eV. 
	
Other possible configurations are
	\begin{equation}
	\label{NisgndCSstates}
	\begin{array}{cccccccc}
	\mathcal{A}\bigl|x_{\uparrow}x_{\downarrow}y_{\uparrow}\bigr\rangle & \!\!\!\!,\! & \mathcal{A}\bigl|x_{\uparrow}x_{\downarrow}y_{\downarrow}\bigr\rangle & \!\!\!\!,\! & \mathcal{A}\bigl|x_{\uparrow}x_{\downarrow}z_{\uparrow}\bigr\rangle & \!\!\!\!,\! & \mathcal{A}\bigl|x_{\uparrow}x_{\downarrow}z_{\downarrow}\bigr\rangle & \!\!\!\!,\!\!\\
	\mathcal{A}\bigl|y_{\uparrow}y_{\downarrow}z_{\uparrow}\bigr\rangle & \!\!\!\!,\! & \mathcal{A}\bigl|y_{\uparrow}y_{\downarrow}z_{\downarrow}\bigr\rangle & \!\!\!\!,\! & \mathcal{A}\bigl|y_{\uparrow}y_{\downarrow}x_{\uparrow}\bigr\rangle & \!\!\!\!,\! & \mathcal{A}\bigl|y_{\uparrow}y_{\downarrow}x_{\downarrow}\bigr\rangle & \!\!\!\!,\!\!\\
	\mathcal{A}\bigl|z_{\uparrow}z_{\downarrow}x_{\uparrow}\bigr\rangle & \!\!\!\!,\! & \mathcal{A}\bigl|z_{\uparrow}z_{\downarrow}x_{\downarrow}\bigr\rangle & \!\!\!\!,\! & \mathcal{A}\bigl|z_{\uparrow}z_{\downarrow}y_{\uparrow}\bigr\rangle & \!\!\!\!,\! & \mathcal{A}\bigl|z_{\uparrow}z_{\downarrow}y_{\downarrow}\bigr\rangle & \!\!
	\end{array}
	\end{equation}
	which reduces into a $|^{2}T_{1}\rangle$ spin doublet as 
	\begin{equation}
	\label{Nisgnd2T1}
	|^{2}T_{1}\rangle=\left\{ \begin{array}{c}
	\mathcal{A}\bigl|x_{\sigma}y_{\downarrow}y_{\uparrow}\bigr\rangle-\mathcal{A}\bigl|x_{\sigma}z_{\downarrow}z_{\uparrow}\bigr\rangle\\
	\mathcal{A}\bigl|y_{\sigma}z_{\downarrow}z_{\uparrow}\bigr\rangle-\mathcal{A}\bigl|y_{\sigma}x_{\downarrow}x_{\uparrow}\bigr\rangle\\
	\mathcal{A}\bigl|z_{\sigma}x_{\downarrow}x_{\uparrow}\bigr\rangle-\mathcal{A}\bigl|z_{\sigma}y_{\downarrow}y_{\uparrow}\bigr\rangle
	\end{array}\right\} \otimes\left\{ \begin{array}{c}
	\sigma=\uparrow\\
	\sigma=\downarrow
	\end{array}\right.\!\!
	\end{equation}
	and an another $|^{2}T_{2}\rangle$ spin doublet as					
	\begin{equation}
	\label{Nisgnd2T2}
	|^{2}T_{2}\rangle=\left\{ \begin{array}{c}
	\mathcal{A}\bigl|x_{\sigma}y_{\downarrow}y_{\uparrow}\bigr\rangle+\mathcal{A}\bigl|x_{\sigma}z_{\downarrow}z_{\uparrow}\bigr\rangle\\
	\mathcal{A}\bigl|y_{\sigma}z_{\downarrow}z_{\uparrow}\bigr\rangle+\mathcal{A}\bigl|y_{\sigma}x_{\downarrow}x_{\uparrow}\bigr\rangle\\
	\mathcal{A}\bigl|z_{\sigma}x_{\downarrow}x_{\uparrow}\bigr\rangle+\mathcal{A}\bigl|z_{\sigma}y_{\downarrow}y_{\uparrow}\bigr\rangle
	\end{array}\right\} \otimes\left\{ \begin{array}{c}
	\sigma=\uparrow\\
	\sigma=\downarrow
	\end{array}\right.\text{.}
	\end{equation}
	
We expect that the levels of these states lie above those of the maximally open shell configurations' doublets because the Coloumb repulsion is not compensated by exchange interaction in the former. In Kohn-Sham DFT method, we could calculate one of the single Slater-determinants electronic configuration as expressed in Eq.~\eqref{NisgndCSstates}. Their total energy exceeds that of $|^{2}E\rangle$ by about 0.5~eV within $T_d$ symmetry.

Finally, we conclude based on the Coulombic repulsion principle that $|^{2}T_{2}\rangle$ has the highest total energy among the considered doublet manifolds as it exhibits the most symmetric polynomial, that is, $x(y^{2}+z^{2})$, in contrast to $|^{2}T_{1}\rangle$ with $x(y^{2}-z^{2})$.

\subsection{$t_{2}^{(2)}$ electronic configurations in Ni$_\text{s}^0$} \label{sec:Nisgndc0}

The ground state manifolds of Ni$_\text{s}^0$ exhibit $t_{2}^{(2)}$ electronic configurations in $T_d$ symmetry which may be a subject of Jahn-Teller effect and spin-orbit coupling. The total number of manifolds is 
$\binom{6}{2}=15$. According to direct product tables~\cite{atkins1970tables} of $T_{d}$ point group, there are four multiplets that can be read as
		\begin{equation}
			\label{Nisc0twoparticle}
	t_{2}^{(2)}={}^{1}A_{1}\oplus{}^{1}E\oplus{}^{3}T_{1}\oplus{}^{1}T_{2} \text{,}
		\end{equation}
where we kept the fermionic anti-symmetric property of the wavefunctions, thus $^{3}A_{1},{}^{3}E,{}^{1}T_{1},{}^{3}T_{2}$ combinations do not occur. Next, we show the symmetrically adapted wavefunctions. According to our \textit{ab initio} results the ground state of $\mathrm{Ni_s^0}$ is a $|^{3}T_{1}\rangle$ that reads as
		\begin{equation}
		\label{Nisc0_3T1_wave}
		|^{3}T_{1}\rangle=\left\{ \begin{array}{c}
		\mathcal{A}\left|yz\right\rangle \\
		\mathcal{A}\left|zx\right\rangle \\
		\mathcal{A}\left|xy\right\rangle 
		\end{array}\right\} \otimes\left\{ \begin{array}{c}
		\left|\uparrow\uparrow\right\rangle \\
		\mathcal{S}\left|\uparrow\downarrow\right\rangle \\
		\left|\downarrow\downarrow\right\rangle 
		\end{array}\right.  \text{.}
		\end{equation}	
It is followed by the $|^{1}T_{1}\rangle$ singlet state which can be expressed as 
	\begin{equation}
	\label{Nisc0_1T2_wave}
	|^{1}T_{2}\rangle=\left\{ \begin{array}{c}
	\mathcal{S}\left|yz\right\rangle \\
	\mathcal{S}\left|zx\right\rangle \\
	\mathcal{S}\left|xy\right\rangle 
	\end{array}\right\} \otimes\mathcal{A}\left|\uparrow\downarrow\right\rangle 
	\text{,}
	\end{equation}	
	where we interchanged the two-particle $\mathcal{A}\left|ab\right\rangle =\frac{1}{\sqrt{2}}(\left|ab\right\rangle -\left|ba\right\rangle )$ anti-symmetrisation and $\mathcal{S}\left|ab\right\rangle =\frac{1}{\sqrt{2}}(\left|ab\right\rangle +\left|ba\right\rangle )$ symmetrisation operators on the spin and orbital part of the wavefunctions.

The triplet configuration can be determined by Kohn-Sham DFT as a $|x^{\uparrow}y^{\uparrow}\rangle$ configuration. We note that $|^{3}T_{1}\rangle$ triplet is Jahn-Teller unstable. We obtained 159~meV relaxation energy when we removed the $T_d$ symmetry constraint during the geometry optimization. We obtained 47~meV higher total energy for the $|x^{\uparrow}y^{\downarrow}\rangle$ with $T_d$ symmetry restriction that gained 125~meV relaxation energy upon removal the symmetry constraint. We note that the $|x^{\uparrow}y^{\downarrow}\rangle$ configuration is the 50\%-50\% admixture of the $|^{3}T_{1}\rangle$ and $|^{1}T_{2}\rangle$ multiplets, thus our results can be interpreted as that the singlets are above the triplet by about $\sim$47~meV.
	
Eqs.~\eqref{Nisc0_3T1_wave} and \eqref{Nisc0_1T2_wave} span a $9+3=12$ dimensional open-shell subspace where the electrons are placed to different orbitals within $t_2$ orbitals. Next, we discuss the case of the closed shell configurations ($xx$, $yy$, $zz$) which leads to the $|^{1}A_{1}\rangle$ and ${|}^{1}E\rangle$ multiplets which may read as
	\begin{equation}
		\label{Nisc0_1E_wave}
		|^{1}E\rangle=\left\{ \begin{array}{c}
		\frac{1}{\sqrt{6}}\bigl(2\left|zz\right\rangle -\left|xx\right\rangle -\left|yy\right\rangle \bigr)\\
		\frac{1}{\sqrt{2}}\bigl(\left|xx\right\rangle -\left|yy\right\rangle \bigr)
		\end{array}\right\} \otimes\mathcal{A}\left|\uparrow\downarrow\right\rangle 
	\end{equation}
	and 
	\begin{equation}
		\label{Nisc0_1A1_wave}
		|^{1}A_{1}\rangle=\frac{1}{\sqrt{3}}\bigl(\left|zz\right\rangle +\left|xx\right\rangle +\left|yy\right\rangle \bigr)\otimes\mathcal{A}\left|\uparrow\downarrow\right\rangle 
		\text{,}
	\end{equation}
respectively. We note that $|^{1}E\rangle$ and $|^{1}A_{1}\rangle$ are multi-configurational states, thus Kohn-Sham DFT may not provide good estimates for their total energies. Nevertheless, we can determine the total energy within spin-averaged Kohn-Sham DFT calculation for the individual ($xx$, $yy$, $zz$) configurations that forces the same orbitals in each spin channel. According to our results, the total energy of the closed shell configurations lies above that of the ground state by 0.70~eV within $T_d$ symmetry. After removing the symmetry constraint in the geometry optimization procedure, the calculated energy difference is reduced to 0.18~eV that shows a giant 0.52~eV Jahn-Teller energy. This effect might suppress the electronic correlation energy between $|^{1}E\rangle$ and $|^{1}A_{1}\rangle$, similarly to the product Jahn-Teller effect in SiV$^0$ defect of diamond~\cite{Thiering2019prodJT, Harris2019}. In the next section, we discuss the Jahn-Teller effect in detail for $t_2^{(2)}$ electronic configuration in $T_d$ symmetry.

\subsubsection{$T\otimes (t \oplus e)$ Jahn-Teller effect in Ni$_\text{s}^0$ \label{sec:Nis0JT}}

\begin{figure*}[] 
	\includegraphics[width=\textwidth]{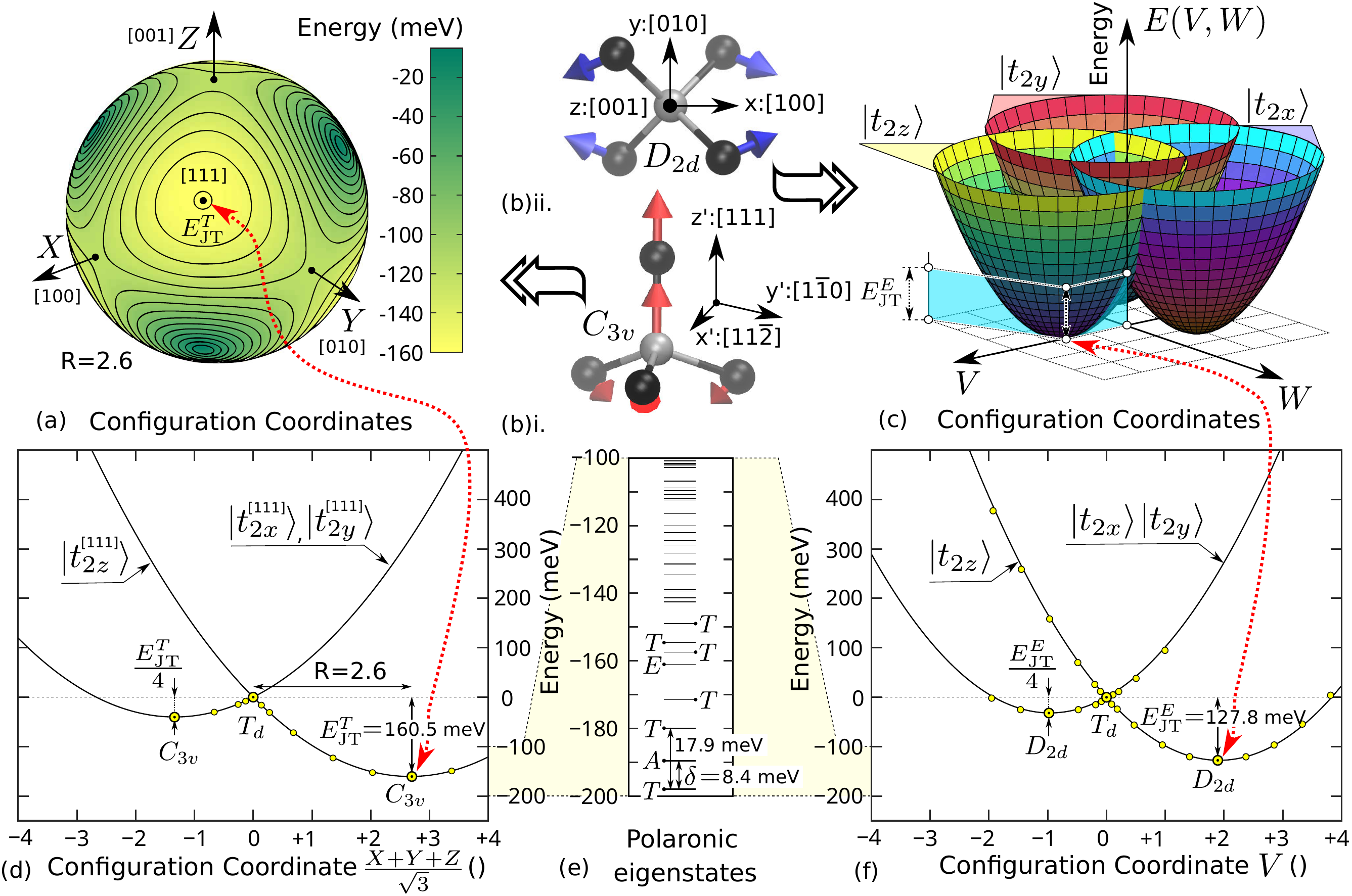} 
	\caption{\label{fig:PESc0fig} (a) Lowest layer of APES for $T$ vibration mode showing the four equivalent minima. (b) Geometry distortions for (i) $E$ and (ii) $T$  type modes towards configuration coordinates $V$ and $(X+Y+Z)/\sqrt(3)$ respectively. (c) APES for $E$ vibration mode depicting the three equivalent minima points towards directions $+V$ and $-\frac{1}{2}V\pm\frac{\sqrt{3}}{2}W$. (d) APES for $T$ vibration mode towards the [111] distortion. Quantization axis and shape for $|t_{2...}^{[111]}\rangle$ orbitals are shown in the second row of Fig.~\ref{fig:NisKS}(b). (e) Polaronic eigenspectrum by means of Eq.~\eqref{Nisc0_JT} for Ni$_\text{s}^0$ without the $\frac{3}{2}\hbar\omega_{T}+\frac{2}{2}\hbar\omega_{E}$ zero point energy terms. (f) APES for $E$ vibration mode towards the $V$ distortion. Quantization axis and shape for $|t_{2...}\rangle$ orbitals are shown in the first row of Fig.~\ref{fig:NisKS}(b). We note that yellow filled-in data points depict \textit{ab inito} results.}
\end{figure*}

The theory of Jahn-Teller interaction in $T_d$ symmetry was already discussed in the literature (see Ref.~\onlinecite{bersuker2013jahn} and references therein). We apply this theory to the Ni$_\text{s}^0$ defect. The $T_d$ symmetry is dynamically distorted by $t$ and $e$ vibration modes, i.e. $T\otimes (t \oplus e)$ vibronic system. With assuming an effective $t$ vibration and an effective $e$ vibration, the vibronic system may read as  
\begin{equation}
	  		\label{Nisc0_JT}
		  	\begin{split}\hat{H}_{\mathrm{JT}}= \hat{H}_{\mathrm{JT}} ^{T} + \hat{H}_{\mathrm{JT}} ^{E} = \\
		  	\hbar\omega_{T}\sum_{n=X,Y,Z}\left(a_{n}^{\dagger}a_{n}+\frac{1}{2}\right)+\hbar\omega_{E}\sum_{n=V,W}\left(a_{n}^{\dagger}a_{n}+\frac{1}{2}\right)\\
		  	+\left[\begin{array}{ccc}
		  	F_{E}\left(\frac{1}{2}\hat{V}-\frac{\sqrt{3}}{2}\hat{W}\right) & -F_{T}\hat{Z} & -F_{T}\hat{Y}\\
		  	-F_{T}\hat{Z} & F_{E}\left(\frac{1}{2}\hat{V}+\frac{\sqrt{3}}{2}\hat{W}\right) & -F_{T}\hat{X}\\
		  	-F_{T}\hat{Y} & -F_{T}\hat{X} & -F_{E}\hat{V}
		  	\end{array}\right] \text{,}
		  	\end{split}
		\end{equation}
		where $\hat{X},\hat{Y},\hat{Z}$ operators depict the geometry distortion from the $t$ vibration mode governed by $\hat{H}_{\mathrm{JT}} ^{T}$ while $\hat{V},\hat{W}$ depict to that of $e$ vibration mode governed by $\hat{H}_{\mathrm{JT}} ^{E}$. $F_E$ and $F_T$ electron-vibration coupling parameters can be determined~\cite{bersuker2013jahn} as
		\begin{equation}
			F_{T}=\sqrt{\frac{3}{2}E_{\mathrm{JT}}^{T}\hbar\omega_{T}} \quad \text{and} \quad
			F_{E}=\sqrt{2E_{\mathrm{JT}}^{E}\hbar\omega_{E}} 
			\label{straincomponent}   \text{,}
		\end{equation}
where the $E_\mathrm{JT}^{T,E}$ is the Jahn-Teller energy and $\hbar\omega_{T,E}$ is the respective effective phonon  $t$ and $e$  frequencies. The global minima in the five-fold configurational space are either one of the 3 equivalent tetragonal extrema with $D_{2d}$ symmetry or 4 equivalent trigonal extrema with $C_{3v}$ symmetry. 
The actual global energy minima, whether tetragonal or trigonal, depend on the properties of the given system. As the $t$ and $e$ vibrations are orthogonal each other and they lead to trigonal and tetragonal distortions, there individual contributions can be determined by DFT calculations as
	\begin{subequations}
		\begin{alignat}{4}
		E_{\mathrm{JT}}^{T}=E_{tot}^{\mathrm{DFT}}(C_{3v})-E_\mathrm{tot}^{\mathrm{DFT}}(T_{d})  \label{TotenDEFa} \\
		E_{\mathrm{JT}}^{E}=E_{tot}^{\mathrm{DFT}}(D_{2d})-E_\mathrm{tot}^{\mathrm{DFT}}(T_{d}) \label{TotenDEFb} \text{,} 
		\end{alignat} 
		\label{TotenDEF}
	\end{subequations}
where $E_\mathrm{tot}^{\mathrm{DFT}}(\Gamma)$ total energies are obtained from HSE06 DFT calculations with geometry optimization under the $\Gamma$ point group. We determined $\hbar\omega_{T,E}$ by mapping the parabolicity of the adiabatic potential energy surface along the $e$ and $t$ normal coordinates. 
		
We solved the JT problem up to 8-phonon limit by substituting our \textit{ab initio} parameters $E_\mathrm{JT}^{T}=160.5$~meV, $\hbar\omega_{T}=45.0$~meV,
$E_\mathrm{JT}^{E}=127.8$~meV, $\hbar\omega_{E}=69.5$~meV into Eqs.~\eqref{straincomponent} and \eqref{Nisc0_JT}. In accordance to previous results~\cite{bersuker2013jahn}, the lowermost eigenvalue of the electron-vibration or vibronic system is a triple degenerate $T$ level which is followed by an $A$ level by 9.0~meV. $|^{3}T_1\rangle$ also experience spin-orbit effect as the maximum degeneracy in the $T_d$ double group is four-fold. In the next section, we study the spin-orbit interaction of this state.
		
\subsubsection{Spin-orbit fine structure of Ni$_\text{s}^{0}$ \label{sec:Nis0LS}}	

We introduce the spin-orbit operator as follows. The $\hat{L}_{x,y,z}$ operators depict an effective $L=1$ orbital moment within the $t_2$ orbitals. One may note that there are two electrons in the $t_2$ orbital, thus 
				\begin{equation}
				\label{Nisc0_LS}
				\hat{H}_{\mathrm{LS}}=\lambda_0\hat{\boldsymbol{L}}\hat{\boldsymbol{S}}=\lambda\left[\begin{array}{ccc}
				& -i\hat{S}_{z} & -i\hat{S}_{y}\\
				i\hat{S}_{z} &  & -i\hat{S}_{x}\\
				i\hat{S}_{y} & i\hat{S}_{x}
				\end{array}\right] \text{,}
				\end{equation}
		where
			\begin{equation}
				\hat{L}_{x}\!=\!\begin{pmatrix}0\\
				& 0 & \!-i\\
				& i & 0
				\end{pmatrix}\quad\hat{L}_{y}\!=\!\begin{pmatrix}0 &  & \!-i\\
				& 0\\
				i &  & 0
				\end{pmatrix}\quad\hat{L}_{z}\!=\!\begin{pmatrix}0 & \!-i\\
				i & 0\\
				&  & 0
				\end{pmatrix} \text{,}
					\label{orbitalOPS}
				\end{equation}
		and $\hat{S}_{x,y,z}$'s are regular spin operators for a given spin multiplicity. For the triplet spin of Ni$_\text{s}^0$, we use
			\begin{equation}
		\hat{S}_{x}\!=\!\frac{1}{\sqrt{2}}\!\begin{pmatrix} & \!1\\
		1 &  & \!1\\
		& \!1
		\end{pmatrix}\;\hat{S}_{y}\!=\!\frac{1}{\sqrt{2}}\!\begin{pmatrix} & \!\!\!-i\\
		i &  & \!\!\!-i\\
		& \!i
		\end{pmatrix}\;\hat{S}_{z}\!=\!\begin{pmatrix}1\\
		& \!0\\
		&  & \!\!\!-1
		\end{pmatrix}\text{.}
		\label{spinOPS}
		\end{equation}
		The complex $i$ unit in  $\hat{L}_{x,y,z}$ emphasizes that the $t_2$ orbitals lose their real-value character when spin-orbit interaction is considered unlike the case of electron-phonon interaction, e.g. Eq.~\eqref{Nisc0_JT}). For example, if the quantisation axis is along the "z" [001] axis then the eigenfunctions are $|ml=\pm1\rangle=\frac{|t_{2x}\rangle\pm i|t_{2y}\rangle}{\sqrt{2}}$ and $ |ml=0\rangle=|t_{2z}\rangle$ that we visualize in Fig.~\ref{fig:NisKS}(b). 
		
We determined the bare intrinsic spin-orbit strength at single particle level by \textit{ab initio} calculations: $\lambda^{(1)}=\langle t_{2z+}^{\uparrow}|\lambda^{(1)}\hat{\boldsymbol{l}}\hat{\boldsymbol{s}}|t_{2z+}^{\uparrow}\rangle-\langle t_{2z-}^{\uparrow}|\lambda^{(1)}\hat{\boldsymbol{l}}\hat{\boldsymbol{s}}|t_{2z-}^{\uparrow}\rangle=6.78$~meV. The single particle $\hat{\boldsymbol{l}}$ operator can be depicted by the same matrices that of Eq.~\eqref{orbitalOPS} in contrast to a single electron's spin $\hat{\boldsymbol{s}}$ that can be depicted by two-times-two Pauli matrices: $\hat{s}_{x}\!=\!\frac{1}{2}\!\bigl(\begin{smallmatrix} & 1\\
		1
		\end{smallmatrix}\bigr)\;\hat{s}_{y}\!=\!\frac{1}{2}\bigl(\begin{smallmatrix} & \!\!\!-i\\
		i
		\end{smallmatrix}\bigr)\;\!\hat{s}_{z}\!=\!\frac{1}{2}\bigl(\begin{smallmatrix}1\\
		& \!\!\!-1
		\end{smallmatrix}\bigr)$. We note that $\lambda^{(1)}$ acts only between single particle levels but Ni$_\text{s}^{0}$ is effectively a two-electron system, see Eq.~\eqref{Nisc0_3T1_wave}. Therefore, the spin-orbit strength for the two-particle system will be $\lambda^{(1)}\langle t_{2z+}^{\uparrow}t_{2z}^{\uparrow}|\mathcal{A}^{\dagger}\hat{\boldsymbol{l}}\hat{\boldsymbol{s}}\mathcal{A}|t_{2z+}^{\uparrow}t_{2z}^{\uparrow}\rangle=\lambda_0\langle^{3}T_{2}^{(mj=+2)}|\hat{\boldsymbol{L}}\hat{\boldsymbol{S}}|{}^{3}T_{2}^{(mj=+2)}\rangle$. In order to solve the previous equation, we introduce an $1/2$ Clebsch-Gordan coefficient for $\lambda_{0}=\frac{\lambda^{(1)}}{2}$ because only $t_{2z\pm}^{\uparrow}$ posseses both angular and spin moments, so $t_{2z}^{\uparrow}$ exhibiting $ml=0$ remains intact from spin-orbit coupling.
		
However, the intrinsic value of spin-orbit coupling is reduced by the $p_{T_1}$ Ham reduction factor. That is, the $\hat{\boldsymbol{L}}$ orbital operators are mediated by strong Jahn-Teller renormalization thus are subject to the so called Ham reduction factors (see section 5.6 about reduction factors in Ref.~\onlinecite{bersuker2013jahn} for details). In other words, the triply degenerate $t_2$ phonons can entangle the three $|t_{2z\pm}\rangle$, $|t_{2z}\rangle$ orbital characters. That is the ground state that can be depicted in a Born-Oppenheimer basis by ladder operators acting on vibration modes: $|\tilde{t}_{2z+}\rangle=c|t_{2z+}\rangle\otimes|0\rangle+d|t_{2z-}\rangle\otimes a_{X}^{\dagger}|0\rangle + ... $, where we depicted only the first terms in the series expansion. Therefore the electron-phonon entangled $|\tilde{t}_{2z+}\rangle$ will exhibit a reduced spin-orbit strength as the ground state is not a pure $|t_{2z+}\rangle$ electronic character anymore: $|t_{2z-}\rangle$ mixes in. The intrinsic orbital angular momentum will be partially quenched by the $p_{T_1}$ factor: $\hat{\boldsymbol{L}}_{\mathrm{eff}}=p_{T_{1}}\langle\hat{\boldsymbol{L}}\rangle_\text{phonons}$. Thus the observable spin-orbit coupling will be partially quenched too: $\lambda_{\mathrm{eff}}=p_{T_{1}}\lambda_0$. We report the reduction factor as $p_{T_1}=0.0349$ by means of Eq.~\eqref{Nisc0_JT} when we capped the phonon quanta as $n\leq9$. Finally, the $\lambda_\mathrm{eff}$ acting for $|^{4}T_{1}\rangle$'s $L=1$ orbital and $S=1$ spin moments will be
		\begin{equation}
		\label{Nism0SOC}
	\hat{H}_{\mathrm{eff}}=\underset{{{\textstyle =\lambda_{\mathrm{eff}}}}}{\underbrace{\frac{\lambda^{(1)}}{2}p_{T_{1}}}}\hat{\boldsymbol{L}}_{\mathrm{eff}}\hat{\boldsymbol{S}}
		\text{,}
		\end{equation}
		where $\lambda_{\mathrm{eff}} = 0.12$~meV. Our positive $\lambda_{\mathrm{eff}}$ is in agreement with Hund's rules: the lowermost state in a $J=0$ ($A_1$) singlet followed by $J=1$ ($T_1$) at $\lambda_{\mathrm{eff}}$ energy and $J=2$ ($T_2\oplus E$) at $3\lambda_{\mathrm{eff}}$, where we depicted double group representations~\cite{koster1963properties} for the $T_d$ symmetry in $(\dots)$ parentheses. However, higher order spin-orbit terms may dominate over the linear $\lambda_{\mathrm{eff}}$ similarly to that of Ni$_\text{s}^{-}$'s excited state that we will describe in Eq.~\eqref{Nism1SOC} in the next section.
	
	\subsection{Optically excited states of Ni$_\text{s}^{-}$ \label{sec:Nism1ex}}
	
We turn to the interpretation of the optical spectra of Ni-related centres. First, we consider the optically allowed excited states of Ni$_\text{s}^{-}$. To this end, one has to go beyond $t_2^{(3)}$ multiplets because no other quartet state can be formed than the $^4A_2$ ground state. As can be seen in Fig.~\ref{fig:NisKS}(a), a resonant $e$ state occurs in the valence band from which an electron may be promoted to the empty $t_2$ state in the spin minority channel.	This results in $e^{(3)}t_2^{(4)}$ electron configuration which is equivalent to $e^{(1)}t_2^{(2)}$ hole configuration. We assume that the $t_2^{(2)}$ prefers the $|^{3}T_{1}\rangle$ multiplet. With this constraint, the $e^{(1)}t_2^{(2)}$ configuration reads as
\begin{equation}
		\label{Nis_m1ex_grp}
		^{3}T_{1}\otimes^{2}E={}^{4}T_{1}\oplus{}^{4}T_{2}\oplus{}^{2}T_{1}\oplus{}^{2}T_{2} \text{.}
\end{equation}
According to Hund's first rule, we expect that the high-spin quartet states are more stable than the low-spin doublet states. We focus now on the discussion of the optically allowed quartet states. In the construction of the quartet states in Eq.~\eqref{Nis_m1ex_grp} we may start with the doublet $E$ state which can be expressed by ($x^2-y^2$; $2z^2-x^2-y^2$) quadratic polinomials. These polynomials are represented by the $|v\rangle=d_{x^{2}-y^{2}}$ and $|w\rangle=d_{2z^{2}-x^{2}-y^{2}}$ Ni $3d$ atomic orbitals as depicted in Fig.~\ref{fig:NisKS}(b). The lowest order polynomial for triple degenerate representation under $T_d$ point group is linear. As a consequence, the quartet states should transform as fourth order polynomials. By applying projection operators in group theory one finds that \{$xy(x^{2}-y^{2})$ ; $zx(z^{2}-x^{2})$ ; $yz(y^{2}-z^{2})$\} belongs to the $T_{2}$ irreducible representation whereas \{$yz(2x^{2}-y^{2}-z^{2})$ ; $zx(2y^{2}-z^{2}-x^{2})$ ; $xy(2z^{2}-x^{2}-y^{2})$\} belongs to the $T_{1}$ representation. These states may be expressed by using $|v\rangle$ and $|w\rangle$ as
\begin{equation}
\label{Nis_m1ex_4T2}
|^{4}T_{2}\rangle=\left.\!\begin{array}{l}
\mathcal{A}\bigl|yz(\frac{1}{2}v+\frac{\sqrt{3}}{2}w)\bigr\rangle\\
\mathcal{A}\bigl|zx(\frac{1}{2}v-\frac{\sqrt{3}}{2}w)\bigr\rangle \\
\mathcal{A}\bigl|xyv\bigr\rangle
\end{array}\right\} \!\otimes\!\left\{ \begin{array}{c}
\left|\uparrow\uparrow\uparrow\right\rangle \\
\!\mathcal{S}\left|\uparrow\uparrow\downarrow\right\rangle \\
\!\mathcal{S}\left|\uparrow\downarrow\downarrow\right\rangle \\
\left|\downarrow\downarrow\downarrow\right\rangle 
\end{array}\right.\!\!\!
		  	\text{,}
		\end{equation}
		and
	\begin{equation}
	\label{Nis_m1ex_4T1}	
|^{4}T_{1}\rangle=\left.\!\!\begin{array}{l}
\mathcal{A}\bigl|yz(\frac{1}{2}w-\frac{\sqrt{3}}{2}v)\bigr\rangle \\
\mathcal{A}\bigl|zx(\frac{1}{2}w+\frac{\sqrt{3}}{2}v)\bigr\rangle \\
\mathcal{A}\bigl|xyw\bigr\rangle
\end{array}\right\} \!\otimes\!\left\{ \begin{array}{c}
\!\!\left|\uparrow\uparrow\uparrow\right\rangle \\
\!\!\mathcal{S}\left|\uparrow\uparrow\downarrow\right\rangle \\
\!\!\mathcal{S}\left|\uparrow\downarrow\downarrow\right\rangle \\
\!\!\left|\downarrow\downarrow\downarrow\right\rangle 
\end{array}\right.\!\!\!
		\text{.}
	\end{equation}		
It may be recognized that $^4T_1$ states can be constructed by swapping $|v\rangle$ and $w\rangle$ in the multiplets of $^4T_2$ states. Another important and surprising observation is that both $^4T_2$ and $^4T_1$ states may be described as a single Slater-determinant (see the last row in Eqs.~\eqref{Nis_m1ex_4T2} and \eqref{Nis_m1ex_4T1}). As a consequence, the total energy of $^4T_2$ and $^4T_1$ states can be principally estimated from KS DFT $\Delta$SCF method once the respective $|xyv\rangle$ and $|xyw\rangle$ electronic configurations could be converged in this procedure. We obtained 2.61~eV and 2.95~eV excitation energies w.r.t.\ the $^4A_2$ ground state's energy within $T_d$ symmetry. 

The electronic configurations and their energy levels may then be expressed by omitting the spin degrees of freedom as follows,
	\begin{equation}
	\label{T1_T2Correlation}
	\hat{W}_{e}=(\Xi+\Lambda)|{}^{4}T_{1}\rangle\langle{}^{4}T_{1}|+\Xi|{}^{4}T_{2}\rangle\langle{}^{4}T_{2}|+ 0|{}^4A_2\rangle\langle{}^4 A_2| \text{,}
	\end{equation}
where we define the $\hat{P}_{2}$ and $\hat{P}_{1}$ projectors as
		\begin{subequations}\label{6times6_projectors}
				\begin{equation}
	\label{6times6_projectorsT2}
\underset{{\textstyle \hat{P}_{2}}}{\underbrace{|{}^{4}T_{2}\rangle\langle{}^{4}T_{2}|}}=\frac{1}{4}\!\left(\begin{array}{cccccc}
1 & \!\!\!-\sqrt{3}\\
\!\!\!-\sqrt{3} & 3\\
&  & 1 & \!\sqrt{3}\\
&  & \!\sqrt{3} & 3\\
&  &  &  & 4 & 0\\
&  &  &  & 0 & 0
\end{array}\right)\!\!\begin{array}{c}
\leftarrow\mathcal{A}\bigl|yzv\rangle\\
\leftarrow\mathcal{A}\bigl|yzw\rangle\\
\leftarrow\mathcal{A}\bigl|zxv\rangle\\
\leftarrow\mathcal{A}\bigl|zxw\rangle\\
\leftarrow\mathcal{A}\bigl|xyv\rangle\\
\leftarrow\mathcal{A}\bigl|xyw\rangle
\end{array}
\end{equation}
and
\begin{equation}
		\label{6times6_projectorsT1}
\underset{{\textstyle \hat{P}_{1}}}{\underbrace{|{}^{4}T_{1}\rangle\langle{}^{4}T_{1}|}}=\frac{1}{4}\!\left(\begin{array}{cccccc}
3 & \!\sqrt{3}\\
\!\sqrt{3} & 1\\
&  & 3 & \!\!\!-\sqrt{3}\\
&  & \!\!\!-\sqrt{3} & 1\\
&  &  &  & 0 & 0\\
&  &  &  & 0 & 4
\end{array}\right)\!\!\begin{array}{c}
\leftarrow\mathcal{A}\bigl|yzv\rangle\\
\leftarrow\mathcal{A}\bigl|yzw\rangle\\
\leftarrow\mathcal{A}\bigl|zxv\rangle\\
\leftarrow\mathcal{A}\bigl|zxw\rangle\\
\leftarrow\mathcal{A}\bigl|xyv\rangle\\
\leftarrow\mathcal{A}\bigl|xyw\rangle
\end{array} \text{.}
\end{equation}
\end{subequations}

However, we learnt from the $^3T_1$ of Ni$_\text{s}^0$ that is a subject to JT effect. Since $^4T_2$ and $^4T_1$ inherit this electron configuration one can suspect that JT effect is not negligible in $^4T_2$ and $^4T_1$ states of Ni$_\text{s}^-$. Indeed, $\sim$84~meV energy is released once we removed the $T_d$ symmetry constraint in our \textit{ab initio} calculations during the $\Delta$SCF procedure. In the next section, we study the JT effect in the $^4T_2$ and $^4T_1$ states with using $\hat{P}_{2}$ and $\hat{P}_{1}$ projectors in Eqs.~\eqref{6times6_projectorsT2} and \eqref{6times6_projectorsT1}.

\subsubsection{$(T_1 \oplus T_2)\otimes (t \oplus e)$ Jahn-Teller effect \label{sec:Nism1JT}}	
\begin{figure*}[] 
	\includegraphics[width=\textwidth]{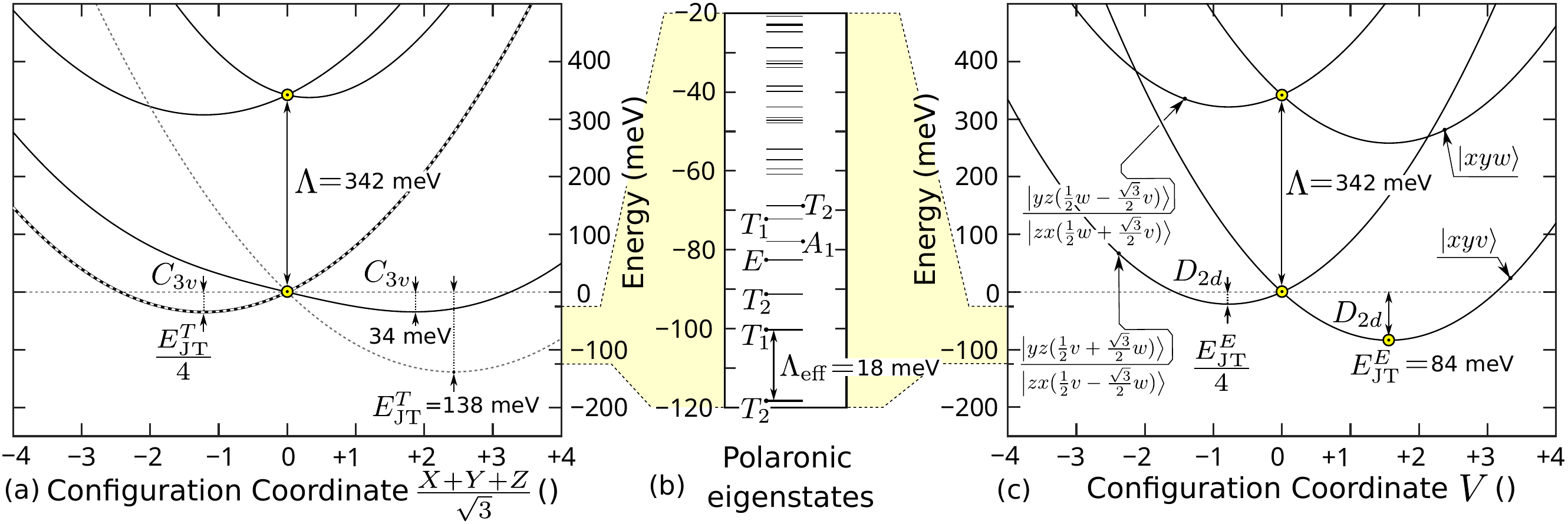} 
	\caption{\label{fig:PESm1exfig} (a) APES for $T$ vibration modes towards the [111] distortion for Ni$_\text{s}^-$'s excited state. The dotted lines represent the APES where $\Lambda =0$ is assumed in Eq.~\eqref{T1_T2Correlation}. It can be seen that $E_{\text{JT}}^{T}/4$ minima at negative distortion is independent on the choice of $\Lambda$. Theoretically, it is possible to determine $E_{\text{JT}}^{T}$ directly by \textit{ab initio} calculations; however, we were not able obtain convergent results. Therefore, we assumed $E_{\text{JT}}^{T}=138$~meV from Ni$_\text{s}^2-$ charge state that exhibits the same electron occupancy in the $t_2$ orbital. We note that yellow filled-in data points are \textit{ab inito} results. (b) Polaronic eigenspectrum by means of Eq.~\eqref{Nism1JT} for Ni$_\text{s}^-$ without the $\frac{3}{2}\hbar\omega_{T}+\frac{2}{2}\hbar\omega_{E}$ zero point energy terms. (c) APES for $E$ vibration modes towards the $V$ distortion. Quantization axis and shape for $t_{2x,y,z}=x,y,z$ orbitals are shown in the first row of Fig.~\ref{fig:NisKS}(b). }
\end{figure*}

We already analysed the JT effect for the $T \otimes (t \oplus e)$ system in Sec.~\ref{sec:Nis0JT}. We rely on this analysis which becomes more complex with taking both $T_2$ and $T_1$ states that might also interact with each other through electron-phonon coupling that we describe by JT effect. This may be described as
\begin{equation}
			\label{Nism1JT}
		\begin{aligned}\hat{H}=\hat{H}_{\mathrm{JT}}+\hat{W}_{e}=\hat{I}\hat{H}_{\mathrm{JT}}\hat{I}+\hat{W}_{e}=& \\
		\hat{P}_{1}\hat{H}_{\mathrm{JT}}^{T}\hat{P}_{1}\!+\!\hat{P}_{1}\hat{H}_{\mathrm{JT}}^{T}\hat{P}_{2}\!+\!\hat{P}_{2}\hat{H}_{\mathrm{JT}}^{T}\hat{P}_{1}\!+\!\hat{P}_{2}\hat{H}_{\mathrm{JT}}^{T}\hat{P}_{2}& + \\
		\underset{\text{diagonal}\,\neq\,0}{\underbrace{\hat{P}_{1}\hat{H}_{\mathrm{JT}}^{E}\hat{P}_{1}}}\!+\!\underset{{\text{offdiagonal\:terms}}\,=\,0}{\underbrace{\xcancel{\hat{P}_{1}\hat{H}_{\mathrm{JT}}^{E}\hat{P}_{2}\!+\!\hat{P}_{2}\hat{H}_{\mathrm{JT}}^{E}\hat{P}_{1}}}}\!+\!\underset{\text{diagonal\,}\neq\,0}{\underbrace{\hat{P}_{2}\hat{H}_{\mathrm{JT}}^{E}\hat{P}_{2}}}&\!+\!\hat{W}_{e}
		\end{aligned}
		 \text{,}
\end{equation}		
where we assume that $\hat{H}_{\mathrm{JT}}$ should not depend on $|v\rangle$ and $|w\rangle$ orbitals. That is, $\hat{H}_{\mathrm{JT}}$ in Eq.~\eqref{Nism1JT} is the dot product of $\hat{H}_{\mathrm{JT}}$ from Eq.~\eqref{Nisc0_JT} with a 2$\times$2 unit matrix to span the six dimensional degrees of freedom from the orbitals. In the derivation, we used the lemma $\hat{I} = \hat{P_1} + \hat{P_2}$ which leads to diagonal and offdiagonal terms where we can distribute the JT interaction acting on inside the $^{4}T_{1,2}$ manifolds (\textit{diagonal terms}) or induce orbital excitation between the $^{4}T_{1,2}$ multiplets. As a consequence, JT interaction induces mixing of electronic characters.

First, one may check that the "$E$" vibration modes do not lead to this kind of orbital mixing that we depicted in Eq.~\eqref{Nism1JT}. That is, Eqs. \eqref{6times6_projectorsT2} and \eqref{6times6_projectorsT1} show that the projectors are diagonal at their bottom-right edges, i.e. $\bigl(\begin{smallmatrix}0\\ & 4 \end{smallmatrix}\bigr)$, $\bigl(\begin{smallmatrix}4\\	& 0 \end{smallmatrix}\bigr)$, respectively. This bottom-right edge is influenced only by the $-F_E \hat{V}$ diagonal matrix element from Eq.~\eqref{Nisc0_JT} and thus does not inflict any interaction between $\mathcal{A}|xyv\rangle$ and $\mathcal{A}|xyw\rangle$ configurations of $|^{4}T_{2}\rangle$ and $|^{4}T_{1}\rangle$ multiplets, respectively. One may argue that the upper two 2$\times$2 matrix blocks in Eq. \eqref{6times6_projectors} are not diagonal. However, these matrix blocks can be diagonalized with $\mathcal{A}|xy(\frac{w}{2}\pm\frac{\sqrt{3}}{2}v)\rangle$, $\mathcal{A}|xy(\frac{v}{2}\pm\frac{\sqrt{3}}{2}w)\rangle$ orbital rotations. Therefore, offdiagonal terms in Eq.~\eqref{Nism1JT} do not appear for $E$-type distortions. Therefore, APES remains the same for $|^{4}T_{1,2}\rangle$ states. They are just shifted by $\Xi$ electronic splitting of Eq.~\eqref{T1_T2Correlation} and thus Eq.~\eqref{TotenDEFb} can be used to determine $F_E$ strength. According to our \textit{ab initio} calculations, the JT distortion energy is $E_\mathrm{JT}^E=84.2$~meV mediated by $\hbar\omega_E=68.7$~meV vibrations.

However, $F_T$ in Eq.~\eqref{TotenDEFa} is effective in this regard. $\hat{X}$, $\hat{Y}$, $\hat{Z}$ in Eq.~\eqref{Nisc0_JT} are purely offdiagonals that can create cross-talks between 2$\times$2 blocks of the projectors in Eq.~\eqref{6times6_projectors} and thus mix the $|^{4}T_{1,2}\rangle$ electronic characters since the three 2$\times$2 blocks are not diagonal in the same basis. As we saw in the previous paragraph, the third block is diagonal in $\{v;w\}$ basis but the first two blocks are diagonal only in $\{ \frac{v}{2}\pm\frac{\sqrt3}{2}w,\frac{w}{2}\mp \frac{\sqrt3}{2} {v} \}$, thus $\hat{H}_\mathrm{JT}^T$ will cause orbital excitation between $|^{4}T_{1,2}\rangle$ multiplets. In other words, the APES will not be just a shifted replicas that of $\mathrm{Ni_s^{0}}$'s $T\otimes (t \oplus e)$ JT APES depicted in Sec.~\ref{sec:Nis0JT}. Indeed, we report that the distortion energy is severely quenched for "$T$"-type distortions: $E_\mathrm{JT}^T=11.6$~meV (that was 160~meV for $\mathrm{Ni_s^{0}}$) that demonstrates that Eq.~\eqref{straincomponent} should not be directly employed for $\mathrm{Ni_s^{-}}$'s excited states. Therefore we used $E_\mathrm{JT}^T=138.4$~meV and $\hbar \omega_T=46.8$~meV that of $\mathrm{Ni_s^{2-}}$ that resembles this excited state. Both configurations confine four electrons in $t_2$ orbitals thus we expect similar JT instability for these two cases approximately. We also assume that the pure atomic $d$ orbitals in $e$ states do not affect the Jahn-Teller instability too much. 

Therefore, we take an exact diagonalisation of Eq.~\eqref{Nism1JT} in order to approximate the vibronic levels up to $n\leq8$ phonon limit. By considering the offdiagonal terms, the $|{}^{4}T_{2}\rangle$ and $|{}^{4}T_{1}\rangle$ can be entangled by phonons which will ultimately lead to a vibronically strongly coupled lowermost $|^{4}\tilde{T}_{2}\rangle$ polaronic state quickly followed by an additional $|^{4}\tilde{T}_{1}\rangle$ state at $\Lambda_{\mathrm{eff}}= 18.0$~meV. We observe that at any choice for $\Lambda$ below $\leq342$~meV the effective energy spacing will be strongly quenched as $\Lambda_{\mathrm{eff}} = p_E \Lambda$ by the $p_E$ Ham reduction factor. We note that $p_E$ affects orbital operators transforming as $E$ representation of $T_d$ symmetry (see section 5.6 about reduction factors in Ref.~\onlinecite{bersuker2013jahn} for details). In our present case, $\Lambda$ energy exhibits the same order of magnitude that of $E_\mathrm{JT}^T$ or $E_\mathrm{JT}^E$ thus the two effects will compete with each other. As a consequence, $p_E$ will be dependent on $\Lambda$ thus it is $p_E(\Lambda=342\:\text{meV})=\frac{18.02\:\text{meV}}{342\:\text{meV}}=0.0527$ in our case.

\subsubsection{Spin-orbit fine structure \label{sec:Nism1LS}}	
We learnt from $^3T_1$ ground state of Ni$_\text{s}^0$ that spin-orbit interaction occurs. Nazar\'e \textit{et al.} was able to fit the following spin-orbit Hamiltonian (see Eq.~(2) in Ref.~\onlinecite{Nazar2001W82.51A1T2}) to the fine structure of the 2.51-eV absorption centre,
\begin{equation}
\label{Nism1SOC}
\hat{H}_{\mathrm{LS}}=\underset{{ =\lambda_{\mathrm{eff}}}}{\underbrace{\lambda_0 p_{T_1}}}\hat{\boldsymbol{L}}_\mathrm{eff}\hat{\boldsymbol{S}}+\kappa(\hat{\boldsymbol{L}}_\mathrm{eff}\hat{\boldsymbol{S}})^{2}+\rho\sum_{\alpha}^{\!\!\!\!x,y,z\!\!\!\!}\hat{L}_{\alpha}^{2}\hat{S}_{\alpha}^{2}
\text{,}
\end{equation}		
where the $\hat{\boldsymbol{L}}$ orbital operators are that of Eq.~\eqref{orbitalOPS} and $\hat{\boldsymbol{S}}$ spin operators are extended for quartet ($S=\frac{3}{2}$) multiplicity to depict the spin-orbit interaction similar way to that of Eq.~\eqref{Nisc0_LS} for Ni$_\text{s}^0$. However, in contrast to that of Eq.~\eqref{Nisc0_LS}, Eq.~\eqref{Nism1SOC} is an effective Hamiltonian in which the $\hat{\boldsymbol{L}}$ orbital operators are mediated by strong Jahn-Teller renormalization, i.e. those are subject to the so-called Ham reduction factors. In this case, the reduction factor is $p_{T_1}$ because our three $\hat{\boldsymbol{L}}$ orbital operators transform as the $T_1$ representation. Therefore, the intrinsic orbital angular momentum will be partially quenched by the $p_{T_1}$ factor: $\hat{\boldsymbol{L}}_{\mathrm{eff}}=p_{T_{1}}\langle\hat{\boldsymbol{L}}\rangle_\text{phonons}$ thus the observable spin-orbit coupling will be partially quenched too, i.e. $\lambda_{\mathrm{eff}}=p_{T_{1}}\lambda_0$. We report the reduction factor as $p_{T_1}=0.0604$ by means of Eq.~\eqref{Nism1JT} with using $n\leq7$ for the expression of polaronic states.

The spin-orbit coupling parameter should be also determined. We were not able to converge the excited state together with spin-orbit coupling, thus we determined $\lambda_0$ by taking the value of Ni$_\text{s}^{2-}$ that bears the same number of electrons on its $t_2$ orbital: $\lambda^{(1)}=\langle t_{2z+}^{\uparrow}|\lambda^{(1)}\hat{\boldsymbol{l}}\hat{\boldsymbol{s}}|t_{2z+}^{\uparrow}\rangle-\langle t_{2z-}^{\uparrow}|\lambda^{(1)}\hat{\boldsymbol{l}}\hat{\boldsymbol{s}}|t_{2z-}^{\uparrow}\rangle=-7.18$~meV. We note that $\lambda_{(1)}$, strictly speaking, acts only on single particle KS levels. Therefore, we introduce an additional $1/3$ Clebsch-Gordan coefficient for $\lambda_0$:
$\lambda^{(1)}\langle t_{2z+}^{\uparrow}t_{2z}^{\uparrow}e_{\pm}^{\uparrow}|\mathcal{A}^{\dagger}\hat{\boldsymbol{l}}\hat{\boldsymbol{s}}\mathcal{A}|t_{2z+}^{\uparrow}t_{2z}^{\uparrow}e_{\pm}^{\uparrow}\rangle=\frac{\lambda^{(1)}}{3}\langle^{4}T_{2}^{(mj=+5/2)}|\hat{\boldsymbol{L}}\hat{\boldsymbol{S}}|{}^{4}T_{2}^{(mj=+5/2)}\rangle$. That is, in other words, neither $t_{2z}^{\uparrow}$ or $e_{\pm}^{\uparrow}$ possesses $l_z$ angular momenta, thus only $t_{2z+}^{\uparrow}$ couples with its $ml=1$ moment and $ms=\frac{1}{2}$ spin. In summary, we report $\lambda_{\mathrm{eff}}$ as $\lambda_{\mathrm{eff}}\approx \frac{\lambda^{(1)}}{3}p_{T_1} = -0.145$~meV that is surprisingly near the experimentally observed $-0.163$~meV measured in the 2.51-eV absorption centre by Nazar{\'{e}} \textit{et al.} in Ref.~\onlinecite{Nazar2001W82.51A1T2}.

However, our current level of theory was not able to determine the additional higher order $\kappa$ and $\rho$ parameters in Eq.~\eqref{Nism1SOC}). That is, determining those coefficients requires second order effects that would utilize $|^{2}T_{1}\rangle$, $|^{2}T_{2}\rangle$, $|^{2}E\rangle$ spin doublet states and even the  $|^{4}A_2\rangle$ ground state through $\sim\lambda_{\pm}^{(1)}|t_{2}^{\downarrow}\rangle\langle e_{}^{\uparrow}|$ flipping terms. We report $\lambda_{\pm}\approx 50$~meV from \textit{ab initio} calculations that would scale the $\varrho,\kappa$ parameters by means of second order perturbation theory as 
\begin{equation}
\label{Nism1SOC2ndorder}
\varrho,\kappa\propto\frac{(\lambda_{\pm}/3)^{2}}{E_{\text{tot.}}(^{4}T_{2})-E_{\text{tot.}}(^{2}T_{2})\!}\approx\frac{\!(\frac{50}{3}\:\text{meV)}^{2\!\!}}{1\:\text{eV}}\!=\!0.28\:\text{meV}
\end{equation}
with assuming a heuristic $\sim$1~eV electronic splitting. Our result in Eq.~\eqref{Nism1SOC2ndorder} is in the same order of magnitude that was reported in Ref.~\onlinecite{Nazar2001W82.51A1T2} as $\kappa=-0.326$~meV, $\rho=+0.532$~meV further supporting the association of 2.51-eV absorption centre to the ${}^4A_2\rightarrow{}^4T_1$ optical transition of Ni$_\text{s}^-$.

\subsection{Excited states of Ni$_\text{s}^0$ \label{sec:Nis0exc}}
We continue the investigation with the excited states of $e^{(1)}t_2^{(3)}$ electronic configuration which could be either $t_2^{\uparrow\uparrow\uparrow}$ or $t_2^{\uparrow\uparrow\downarrow}$ spin configurations coupled to $^2E$ which is a hole left on the $d$ orbital with $e$ symmetry with either $\uparrow$ or $\downarrow$ spin state. We start the analysis with the $t_2^{\uparrow\uparrow\uparrow}$ case which reads as 
\begin{equation}
	\label{Nis_c0_exGRP}
	{}^{4}A_{2}\otimes^{2}E={}^{5}E\oplus{}^{3}E \text{.}
\end{equation}
The $|^{3}E\rangle$ is the case when the $S=3/2$ spin from $|^{4}A_{2}\rangle$ and the $S=1/2$ spin from the $|e\rangle$ hole is coupled by the opposite spin states. The maximally spin-polarised wavefunctions are the followings,
\begin{equation}
	\label{Nis_c0_5E}
	\begin{array}{cl}
	|^{5}E\rangle_{ms=+2}= & \mathcal{A}|x_{\uparrow}y_{\uparrow}z_{\uparrow}\xi_{\uparrow}\rangle\\
	|^{3}E\rangle_{ms=+1}= & \frac{1}{\sqrt{12}}\left(3\mathcal{A}|x_{\uparrow}y_{\uparrow}z_{\uparrow}\xi_{\downarrow}\rangle-\mathcal{A}|x_{\uparrow}y_{\uparrow}z_{\downarrow}\xi_{\uparrow}\rangle\!\!\right.\\
	& \left.\!\!-\mathcal{A}|x_{\uparrow}y_{\downarrow}z_{\uparrow}\xi_{\uparrow}\rangle-\mathcal{A}|x_{\downarrow}y_{\uparrow}z_{\uparrow}\xi_{\uparrow}\rangle\right) \text{,}
\end{array}
\end{equation}
where the $|\xi\rangle$ can be either $|v\rangle$ or $|w\rangle$.
We note that the $\mathcal{A}|x_{\uparrow}y_{\uparrow}z_{\uparrow}\xi_{\downarrow}\rangle$ single Slater-determinant dominates ($75\%$) the $|^{3}E\rangle_{ms=+1}$ 	wavefunction, thus we expect that its total energy can be well estimated by $\Delta$SCF DFT method which results in 2.61~eV excitation energy within $T_{d}$ symmetry. The JT energy is negligibly small (3~meV). The total energy of the $^{5}E$ state should be accurate by $\Delta$SCF DFT method as a single Slater-determinant, and it yields 2.08~eV excitation energy. The JT energy is again tiny (5~meV) as expected.
	
We now turn to the other manifolds based on $t_2^{\uparrow\uparrow\downarrow}$ electronic configurations which result in $^{2}E$, $^{2}T_{1}$ and $^{2}T_{2}$ multiplets. We already proved during the discussion of Ni$_\text{s}^-$ that $^{2}E$ multiplet exhibits the lowest total energy. Therefore, we only focus to that manifold as it is combined with a hole left on the $d$ orbital with $e$ symmetry which reads as follows
\begin{equation}
	\label{Nis_c0_exGRP2}
	^{2}E\otimes^{2}E={}^{3}A_{1}\oplus{}^{3}A_{2}\oplus{}^{3}E\oplus{}^{1}A_{1}\oplus{}^{1}A_{2}\oplus{}^{1}E \text{.}
\end{equation}
We assume again that the singlet manifolds lie at higher energies than the triplet manifolds. We express the maximal spin states of the optically allowed triplet manifolds,
	\begin{widetext} 	
	\begin{equation}
	\label{Nis_c0_exwave}
\begin{array}{cc}
|^{3}A_{1}\rangle= & \mathcal{A}\left|w_{\uparrow}\frac{1}{\sqrt{12}}\bigl(2x_{\uparrow}y_{\uparrow}z_{\downarrow}-x_{\uparrow}y_{\downarrow}z_{\uparrow}-x_{\downarrow}y_{\uparrow}z_{\uparrow}\bigr)+\frac{1}{2}v_{\uparrow}\bigl(x_{\uparrow}y_{\downarrow}z_{\uparrow}-x_{\downarrow}y_{\uparrow}z_{\uparrow}\bigr)\right\rangle \\
|^{3}E\rangle=\! & \left\{ \begin{array}{c}
\mathcal{A}\left|w_{\uparrow}\frac{1}{\sqrt{12}}\bigl(2x_{\uparrow}y_{\uparrow}z_{\downarrow}-x_{\uparrow}y_{\downarrow}z_{\uparrow}-x_{\downarrow}y_{\uparrow}z_{\uparrow}\bigr)-\frac{1}{2}v_{\uparrow}\bigl(x_{\uparrow}y_{\downarrow}z_{\uparrow}-x_{\downarrow}y_{\uparrow}z_{\uparrow}\bigr)\right\rangle \\
\mathcal{A}\left|w_{\uparrow}\frac{1}{2}\bigl(x_{\uparrow}y_{\downarrow}z_{\uparrow}-x_{\downarrow}y_{\uparrow}z_{\uparrow}\bigr)+\frac{1}{\sqrt{12}}v_{\uparrow}\bigl(2x_{\uparrow}y_{\uparrow}z_{\downarrow}-x_{\uparrow}y_{\downarrow}z_{\uparrow}-x_{\downarrow}y_{\uparrow}z_{\uparrow}\bigr)\right\rangle 
\end{array}\!\right.\\
|^{3}A_{2}\rangle= & \mathcal{A}\left|w_{\uparrow}\frac{1}{2}\bigl(x_{\uparrow}y_{\downarrow}z_{\uparrow}-x_{\downarrow}y_{\uparrow}z_{\uparrow}\bigr)-\frac{1}{\sqrt{12}}v_{\uparrow}\bigl(2x_{\uparrow}y_{\uparrow}z_{\downarrow}-x_{\uparrow}y_{\downarrow}z_{\uparrow}-x_{\downarrow}y_{\uparrow}z_{\uparrow}\bigr)\right\rangle 
\end{array} \text{,}
	\end{equation}
	\end{widetext}
where the other spin states can be readily obtained and explicitly given here. It can be observed that these states are highly correlated. The largest determinant contribution is $\bigl(\frac{2}{\sqrt{12}}\bigr){}^{2}=33\%$ prefactors to where the $\Delta$SCF DFT results can converge to 
\begin{equation}
	\label{Nis_c0_exSingleDet}
	|^{3}A_{1}\rangle\sim|w_{\uparrow}x_{\uparrow}y_{\uparrow}z_{\downarrow}\rangle\:\text{or}\:|^{3}A_{2}\rangle\sim|v_{\uparrow}x_{\uparrow}y_{\uparrow}z_{\downarrow}\rangle\,\text{.}
\end{equation} 	 
As a consequence, the total energies of these individual triplet states cannot be well determined by $\Delta$SCF calculations. Nevertheless, we calculated the $\Delta$SCF total energy of Eq.~\eqref{Nis_c0_exSingleDet}. We obtain 2.45~eV excitation energy within $T_d$ symmetry. The calculated JT energy is 43~meV which is expected due to the partially filled $t_2$ orbitals. As the direct calculation of these states is not feasible with DFT methods, we do not analyze these states further but we give a very rough estimate for their ZPL energies at around 2.4~eV.

\subsection{Ground state of Ni$_\text{s}^+$ \label{sec:Nisp1}}

We briefly discuss paramagnetic Ni$_\text{s}^+$ which may be observed in special p-type doped diamond sample. Ni$_\text{s}^+$ exhibits $t_2^{(1)}$ electronic configuration which yields $|{}^2T_2\rangle$ multiplet. This is a subject to JT effect and spin-orbit interaction.

We can apply JT theory as given in Eqs.~\eqref{Nisc0_JT} and \eqref{straincomponent} which results in $E_\mathrm{JT}^{T} = 177.4$~meV, $\hbar\omega_T=46.8$~meV and $E_\mathrm{JT}^{E} = 115.2$~meV, $\hbar\omega_E=66.7$~meV. These results imply that the Ham reduction factor is significant that may act on the spin-orbit energy gaps. We determined $p_{T_1}=0.0209$ up to the 8-phonon limit. We determined the spin-orbit coupling $\lambda_0$ as 8.6~meV. We note here that no Clebsch-Gordan coefficient is needed in this case since the $|{}^2T_2\rangle$ multiplet can be described as a single $t_{2(\pm/0)}^{\uparrow/\downarrow}\rangle$ orbital. Therefore, the spin-orbit coupling will lead to an $J=1/2$ followed by an high moment $J=3/2$ state with $\lambda_\text{eff}=p_{T_1} \lambda_0= 0.179$~meV. We again note that higher order spin-orbit terms may appear similarly to that of Eq.~\eqref{Nism1SOC}.

\section{Discussion} 
\label{sec:discuss}

\begin{figure*}[] 
			\includegraphics[width=400 pt ]{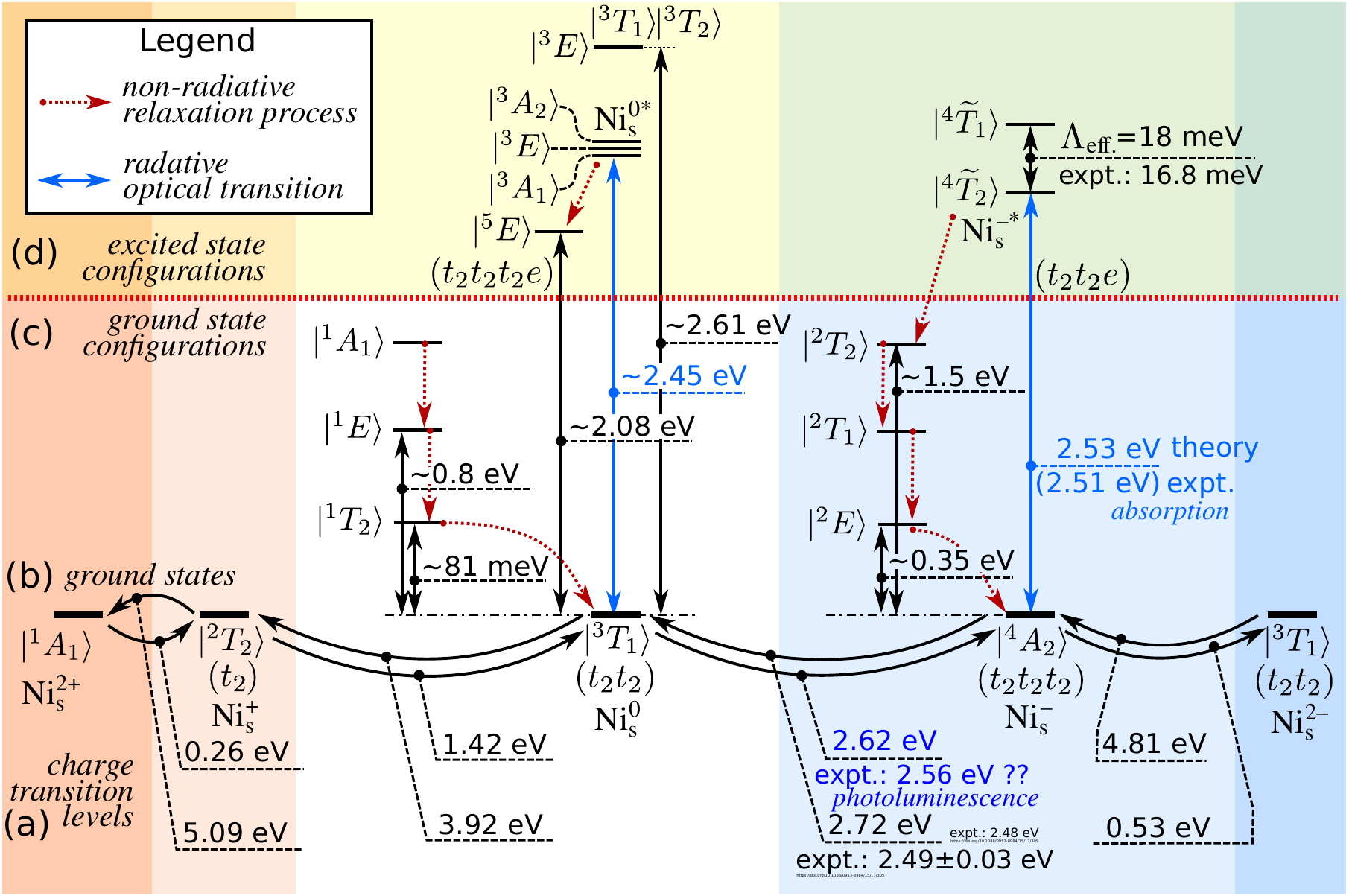} 
			\caption{\label{fig:levels} Ground and excited states of Ni$_\text{s}$ defect in diamond in various charge states. The estimated zero-phonon-line energies of neutral excitation and photo-ionisation threshold energies are also given.}
\end{figure*}

We summarise the calculated excitation energies and photo-ionisation threshold energies of Ni$_\text{s}$ defects in diamond in Fig.~\ref{fig:levels}. As we can see the neutral and negatively charged Ni$_\text{s}$ has photo-ionisation energies and excitation energies are close to 2.5-2.7~eV, thus one has to study the nature of optical excitations in great detail. This includes the vibronic states and levels as well as the spin-orbit couplings and fine structure levels (see Fig.~\ref{fig:levels}). These results are the base for the discussion of the origin of 2.51-eV absorption centre and 2.56-eV ODMR centre. We note that the $\sim$2.5~eV and $\sim$3.0~eV threshold energies in the photo-EPR measurements~\cite{Pereira2003} as the de-activation and re-activation of the W8 EPR centre associated with the ground state of Ni$_\text{s}^-$ cannot be explained by simple ionisation and re-ionisation processes of Ni$_\text{s}^-$, according to our results. In particular, we show below multiple evidences that the 2.51-eV absorption centre is associated with Ni$_\text{s}^-$, thus the photo-ionisation threshold energy towards the conduction band minimum cannot be smaller than 2.51~eV. It is rather likely that carriers trapping is also involved in the process where the carriers might be generated by photo-ionisation of other defects in that diamond sample. 

\subsection{The 2.51-eV absorption centre and Ni$_\text{s}^-$} \label{sec:experimentsm1}
	
We reiterate here that the 2.51-eV absorption centre is associated with a $|{}^4A_2\rangle\rightarrow |{}^4T_2\rangle$ optical transition according to the last observations which shows a replica at 16.5~meV above the ZPL feature~\cite{Nazar2001W82.51A1T2}. Both the ZPL and the replica show a fine structure with about $\sim$1.5~meV splitting where the ZPL's fine structure is depicted by Eq.~\eqref{Nism1SOC}'s effective spin-orbital Hamiltonian. The Ni$_\text{s}^0$ has $T_1$ orbital ground state whereas Ni$_\text{s}^-$ has the appropriate $A_2$ ground state, thus we analyse the excited state of Ni$_\text{s}^-$ as the possible candidate for the 2.51-eV absorption center. Indeed, the calculated ZPL energy is at 2.53~eV, close to the experimental one. Additionally, we were able to predict the experimentally observed SOC parameter $\lambda_\text{expt}=-0.163$~meV as $\lambda_\text{eff}=-0.145$~meV suggesting that the model in Ref.~\onlinecite{Nazar2001W82.51A1T2} is correct.

We found that the lowest energy quartet excited state is indeed $|{}^4\widetilde{T}_2\rangle$ for Ni$_\text{s}^-$ which is optically forbidden in the first order, see Fig.~\ref{fig:PESm1exfig}(b). The next quartet state is a $^4|\widetilde{T}_1\rangle$ separated by about 18.02~meV from the $|{}^4\widetilde{T}_2\rangle$ state that is very close to the observed replica at 16.5~meV. We emphasize that this effect is very similar that we found for the neutral silicon-vacancy center in diamond~\cite{Thiering2019prodJT} where two polaronically entangled states lie close to each other in the Jahn-Teller solution. An additional analog to the properties of the neutral silicon-vacancy optical centre in diamond is that the first state is an optically forbidden $^3|\widetilde{A}_{2u}\rangle$ which is quickly followed by the optically allowed $^3|\widetilde{E}_{u}\rangle$ by $\sim7$~meV that is an order of magnitude smaller than that of the vibration mode's energy ($\hbar\omega=76$~eV). A similar situation is apparent for Ni$_\text{s}^-$: $\Lambda_{\text{expt}}=16.8\:\text{meV}<47\:\text{meV}=\hbar\omega_{T}$. Therefore, the vibration replica is significantly broader ($\sim5$~meV, see Fig. 1 in Ref.~\onlinecite{Nazar2001W82.51A1T2}) than that of the ZPL ($\sim2$~meV) but not as broad as that of a real first phonon vibronic replica ($\sim30$~meV) of Huang-Rhys theory~\cite{Alkauskas:NJP2014, Thiering2017SOC}. Therefore, we argue that the 16.8-meV splitting is a fingerprint of the $(T_1 \oplus T_2)\otimes (t \oplus e)$ Jahn-Teller instability.

The optical activity of the dark $|{}^4\widetilde{T}_2\rangle$ state may be explained by the JT coupling to the bright $|{}^4T_1\rangle$ multiplet which brings a $|{}^4T_1\rangle$ character to the $|{}^4\tilde{T}_2\rangle$ state of about 5\% due to offdiagonal terms in Eq.~\eqref{Nism1JT}. The remaining 95\% of $|{}^4T_1\rangle$'s optical transition dipole moment would be still active around $\Xi+\Lambda=2.51\:\text{eV}+0.34\:\text{eV}=2.85\:\text{eV}$ (435~nm). However, this peak may be not recognized because it is simply obscured by the phonon sideband of the 2.51-eV peak and overlapping with the photo-ionisation threshold at about 2.72~eV converting Ni$_\text{s}^-$ to Ni$_\text{s}^0$. Unfortunately, absorption data beyond $>2.54$~eV ($<488$~nm) is not reported for the 2.51-eV centre, to our best knowledge, to gain insight about this issue. 

Next, we discuss the strength of the ZPL and the replica. First, we note that even the ZPL of the lowermost $|{}^4\widetilde{T}_2\rangle$ only contains 7\% $|{}^4{T}_1\rangle$ contribution and 1\% Debye-Waller factor, thus only 1\% optical transition dipole moment is visible in the replica of 2.51~eV. Additionally, the $\lambda_{\pm}|t_{2}^{\uparrow}\rangle\langle e^{\downarrow}|$ spin-orbit coupling may turn $\langle^{4}A_{2}|\hat{\boldsymbol{d}}|{}^{4}T_{2}\rangle$ optically allowed by $\bigl(\frac{\lambda_{\pm}}{\Xi}\bigr)^{2}\approx\bigl(\frac{50\;\text{meV}}{2510\:\text{meV}}\bigr)^{2}=0.04\%$ amount. 
	
The missing fluorescence of the 2.51-eV band may be explained by the fact that the optical transition dipole moment is weak and $|^2T_2\rangle$, $|^2T_1\rangle$ and $|^2E\rangle$ doublets exist between the 2.51-eV optical transition that could lead to an efficient non-radiative channel for the $^4T_2$ excited state. The estimated spin-orbit coupling between the $^2E$ and $^4A_2$ states is about $\lambda_\pm\sim50$~meV which mediates the intersystem crossing in the non-radiative process.

\subsection{The 2.56-eV ODMR centre and Ni$_\text{s}^0$} \label{sec:experimentsc0}
	
The 2.56-eV ODMR centre is associated with the ground state of Ni$_\text{s}^-$ but observed in luminescence which may be explained by the optical emission from Ni$_\text{s}^0$ as speculated in Ref.~\onlinecite{Nazare1995}. We discuss this scenario in the light of our results. One of the $^3E$ excited states could be calculated with relatively high accuracy at 2.61~eV which is resonant with the calculated photo-ionisation threshold energy at 2.62~eV. The excitation energy of the other three triplet excited states could not well calculated by $\Delta$SCF method and it might fall below the photo-ionisation threshold. However, the strength of optical transition associated with $|t_2\rangle \rightarrow |e\rangle$  at Kohn-Sham orbital level is very weak because the $t_2$ state is well localised on Ni $3d$ orbitals. We think that it is unlikely that those states are involved in the ODMR signal. We here sketch another scenario.

\begin{figure}[] 
	\includegraphics[width=200pt]{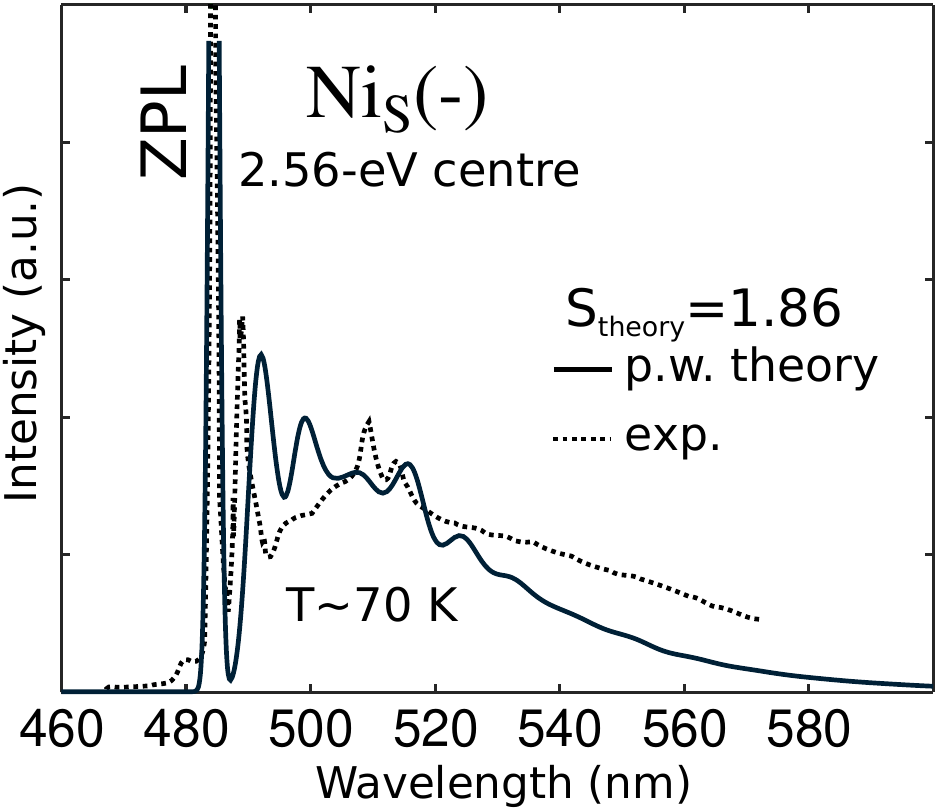} 
	\caption{\label{fig:HR} Huang-Rhys theory for the 2.56-eV PL/ODMR centre (exp.: Ref.~\onlinecite{Nazare1995}). Simulated Ni$_\text{s}^-$+$\text{hole}^+$ bound exciton $\rightarrow$ Ni$_\text{s}^0$ fluorescence process. The calculated Huang-Rhys factor ($S_\text{theory}$) is also given.}
\end{figure}

As we discussed in Sec.~\ref{sec:KS}, Ni$_\text{s}$ spends in the negative charge state upon ultraviolet illumination according to our calculations, and ultraviolet illumination also releases holes in diamond samples~\cite{Nazare1995}. Ni$_\text{s}^-$ can trap holes with very high hole capture rates that results in bound exciton states of Ni$_\text{s}^0$. We note that ODMR signal via bound exciton excited states have been observed for the neutral silicon-vacancy defect~\cite{Zhang2021}, and it was argued for nitrogen-vacancy centre that bound exciton states have giant hole capture cross section~\cite{Lozovoi2021}. We analysed the bound exciton states for silicon-vacancy centre in $D_{3d}$ symmetry~\cite{Zhang2021}. For Ni$_\text{s}^-$ plus hole system, we should consider $T_d$ symmetry as the Ni$_\text{s}^-$ ground state electronic configuration is stable in $T_d$ symmetry and the hole split from valence band maximum should have negligible JT effect. For the hole state, we may use the effective mass theory as explained in Ref.~\onlinecite{Zhang2021}. The lowest energy state transforms by $A_1$ symmetry which has an $1s$ type envelope function. The $1s$ effective mass state is relatively well localised around the core of the defect for which the optical transition dipole moment is relatively strong. The phonon sideband of the fluorescence spectrum may be well estimated by the optimized geometries of the Ni$_\text{s}^-$ representing the excited state and Ni$_\text{s}^0$ for the ground state within Franck-Condon approximation. Indeed, the calculated and observed PL spectra agree well (see Fig.~\ref{fig:HR}). We conclude that the 2.56-eV PL and ODMR signals are associated with the bound exciton emission of Ni$_\text{s}^0$.

Understanding the ODMR contrast requires the analysis of the fine structure of the ground state and the excited state. The ground state is $^3T_1$ which splits to $J=0$ ($A_1$) state, $J=1$ ($T_1$) state and $J=2$ ($T_2 \oplus  E$) states in ascending energy order where we label here the double group representations. The splitting is caused by the spin-orbit interaction which is roughly estimated to be at 0.12~meV based on our \textit{ab initio} calculations that do not contain quadratic spin-orbit contributions that might be also significant. Nevertheless, we assume that the order of these levels retains. In this case, the ground state has no effective coupling to the external magnetic field in the ground state, thus it may be not observable in EPR measurements. The Jahn-Teller nature of the ground state could also make challenging to observe the triplet state in EPR experiments. In the excited state, the $^4A_2$ electronic configuration has $S_1=3/2$ spin state which is combined with a hole in the valence band maximum with $S_2=1/2$ state ($T_2$). The composed system transforms at $T_1$ which again results in $S=0$ $A_1$ state, $S=1$ $T_1$ state and $S=2$ $T_2 \oplus  E$ states split by spin-orbit interaction, where $S=S_1+S_2$ here. The initial population of the excited state branch may depend on the applied external magnetic field, temperature and strain in the sample. The final strength and sign of the ODMR contrast will depend on the population of these states in the excited state manifold. Further analysis of this process and the $g$-factor of the state is out of the scope of the present study.

\subsection{Interaction of nickel defects with other defects in diamond} \label{sec:defectscomparison}

In the previous sections we already outlined that defect-defect interactions play a crucial role in the manifestation of the ODMR signal associated with the substitutional nickel defect in diamond. Since the ionisation energies of defects are much larger than the operation temperatures (typically, room temperature) the exchange of carriers are mediated by photo-excitation. The nitrogen donor can be efficiently activated by photo-excitation energy at 2.1~eV and above. The optical signals of substitional nickel defects have ZPL energies at 2.51~eV and above, thus surrounding nitrogen donors around the nickel defects provide electrons towards the nickel defects upon illumination which may stabilise their negative charge state. On the other hand, if nickel-vacancy (NiV) defects are also present in the sample then this illumination results in ejecting holes by converting the 1.4-eV PL centre (negatively charged NiV defect) to doubly negative charged NiV defects (see Ref.~\cite{Thiering2021}). Thus, the ratio and location of the nitrogen and NiV defects with respect to the substitutional nickel defects can be decisive about the magneto-optical fingerprints of the substitional nickel defects. If vacancies or vacancy aggregates are also present in diamond, e.g. after implantion, then they also enter as traps for holes (in their negative charge states) and as sources of holes upon illumination of neutral vacancies or vacancy clusters. The latter can indeed occur as the calculated acceptor level of the vacancy or divacancy with respect to VBM is lower than 2.5~eV~\cite{Deak2014}. These defects and possibly boron acceptors are the major players in the photodynamics of the substitional nickel defect in diamond.

\section{Summary} \label{sec:summary}
  
We studied the electronic structure of the Ni $3d$ transition metal substitutional defect	with $T_d$ symmetry by means of group theory analysis and plane wave supercell density functional theory calculations. We find that the negative and neutral charge states are the most relevant configurations for the substitutional Ni defect in diamond. We observed strong electron-phonon coupling and spin-orbit coupling in certain states of the defect. We compared the \textit{ab initio} results with the previously reported 2.51-eV absorption centre and the 2.56-eV ODMR centre in diamond, and we associate these centres to the negatively charged defect and emission from the bound exciton excited state of the neutral defect in diamond, respectively.
	
\section*{Acknowledgements}
Support from the National Research Development and Innovation Office of Hungary (NKFIH) within National Excellence Program for the project ``Quantum-coherent materials'' (Grant No.~KKP129866) and the Quantum Information National Laboratory (Grant No.~2022-2.1.1-NL-2022-00004) sponsored by the Cultural and Innovation Ministry of Hungary, from the EU QuantERA II program for the MAESTRO project and the Horizon Europe EIC Pathfinder program for the project QuMicro (Grant No.~101046911) is acknowledged. G.~T.\ acknowledges the support from the J\'anos Bolyai Research Scholarship of the Hungarian Academy of Sciences. We thank the National Information Infrastructure Development Program for the high-performance computing resources in Hungary.


\bibliographystyle{my_custom}
\bibliography{references}
\end{document}